\documentclass[reprint,superscriptaddress,amsmath,amssymb,aps,prx,]{revtex4-2}

\usepackage{graphicx}
\usepackage{bm}
\usepackage{amssymb}
\usepackage{pifont}
\usepackage{xcolor}
\usepackage{orcidlink}

\usepackage{tikz}
\usepackage{tikz-3dplot}
\usepackage{amsmath}

\begin{document}

\title{Alternative LISA-TAIJI networks: polarization separation of the stochastic gravitational wave background}

\author{Gang Wang,  \,\orcidlink{0000-0002-9668-8772}}
\email[Gang Wang: ]{gwanggw@gmail.com, gwang@nbu.edu.cn}
\affiliation{Institute of Fundamental Physics and Quantum Technology, Ningbo University, Ningbo, 315211, China}
\affiliation{Department of Physics, School of Physical Science and Technology, Ningbo University, Ningbo, 315211, China}

\author{Lang~Liu\orcidlink{0000-0002-0297-9633}}
\email[Lang Liu: ]{liulang@bnu.edu.cn}
\affiliation{Department of Physics, Faculty of Arts and Sciences, Beijing Normal University, Zhuhai 519087, China}

\date{\today}

\begin{abstract}

Stochastic gravitational-wave backgrounds (SGWBs) provide a unique opportunity to probe both unresolved astrophysical populations and fundamental physics in the early Universe. Future space-based gravitational-wave (GW) detectors, such as LISA and TAIJI, will enable cross-correlation observations that are particularly sensitive to SGWBs and their polarization content. In this work, we investigate the capabilities of two proposed LISA-TAIJI networks, LISA-TAIJIp and LISA-TAIJIm, for detecting and discriminating tensor, vector, and scalar polarization components of isotropic SGWBs. Using the cross-correlation between the two networks, we evaluate their sensitivities and signal-to-noise ratios for SGWBs containing different combinations of polarization sectors. We find that the LISA-TAIJIm configuration, which has a larger relative inclination between the two constellations, exhibits substantially weaker polarization degeneracies and significantly improved performance in separating polarization components, particularly for tensor-vector-scalar backgrounds. Fisher-matrix forecasts for power-law SGWBs show that the two configurations yield broadly comparable constraints on spectral parameters, reflecting the different roles of component separation and parameter estimation. Our results demonstrate that large-angle LISA-TAIJI networks provide a more favorable geometry for model-independent SGWB polarization measurements.

\end{abstract}

\maketitle

\section{Introduction}

The polarization content of gravitational waves (GWs) provides a direct probe of the fundamental nature of gravity. In general relativity, only two tensor polarization modes are allowed, whereas many alternative theories of gravity predict additional vector and/or scalar polarization states \cite{Brans:1961sx,Damour:1992we,Buchdahl:1970yn,Horndeski:1974wa,Jacobson:2000xp,Bekenstein:2004ne,Hou:2017bqj,Liang:2017ahj,Gong:2018cgj,Dong:2023bgt,Lai:2024fza}. The detection of non-tensorial GW polarizations would provide compelling evidence for physics beyond general relativity \cite{Eardley:1973br,Eardley:1973zuo,Will:2014kxa}. Among the various observational targets, stochastic gravitational-wave backgrounds (SGWBs) are particularly promising because different polarization sectors leave distinct signatures in the cross-correlation signals measured by detector networks \cite{Callister:2017ocg,Nishizawa:2009bf}.

The separation of GW polarization components has been extensively investigated in both ground-based and pulsar-timing observations. For ground-based detectors, the LIGO-Virgo-KAGRA (LVK) collaboration has developed a comprehensive framework for searching for tensor, vector, and scalar components in isotropic SGWBs and has placed upper limits on the corresponding polarization amplitudes using cross-correlation measurements \cite{Nishizawa:2009bf,Callister:2017ocg,LIGOScientific:2018czr,LIGOScientific:2019vic,Abbott:2021ksc,KAGRA:2021kbb,LIGOScientific:2025bgj,LIGOScientific:2025kry}. Similar studies have also been carried out using compact-binary coalescences, where the differing antenna responses of geographically separated detectors provide sensitivity to the polarization content of GWs and enable tests of non-GR polarization states \cite{LIGOScientific:2017ycc,Takeda:2018uai,LIGOScientific:2021sio,LIGOScientific:2026qni,LIGOScientific:2026fcf,Saffer:2020xsw}. At much lower frequencies, pulsar timing array observations have explored the polarization content of nanohertz SGWBs through the characteristic angular correlation patterns induced in pulsar timing residuals \cite{NANOGrav:2023ygs,Chen:2021wdo,Chen:2023uiz}.

Future space-based GW observatories, including LISA, TAIJI, and TianQin, are expected to open the milli-Hz GW window in the mid-2030s \cite{LISA:2017pwj,LISA:2024hlh,Hu:2017mde,TianQin:2015yph}. Their long-duration observations, stable baselines, and high sensitivities make them particularly well suited for SGWB measurements. In addition, networks composed of multiple space-based detectors enable cross-correlation observations analogous to those employed by ground-based interferometers, providing powerful opportunities for detecting and characterizing SGWBs. Previous studies have shown that joint observations by space-based detector networks can significantly enhance the scientific return for massive black hole binaries \cite{Ruan:2020smc,Wang:2020a,Wang:2021polar,Shuman:2021ruh}, SGWB \cite{Omiya:2020fvw,Seto:2020mfd,Orlando:2020oko,Pol:2021uol,Zhao:2024yau}, and cosmological measurements \cite{Wang:2021srv,Wang:2020dkc}. Understanding how the relative geometry of detector constellations influences SGWB polarization measurements is therefore an important problem for the design and optimization of future space-based detector networks.

Three alternative orbital configurations for TAIJI have been proposed to form detector networks with LISA \cite{Wang:2021uih}. Owing to their distinct relative geometries, these configurations exhibit different observational characteristics \cite{Wang:2021uih,Wang:2021njt,Chen:2024fto,Zhang:2026djr}. Previous studies have shown that the LISA-TAIJIm configuration generally provides improved performance for massive black hole binary observations and parity-violating SGWB searches, whereas the nearly colocated LISA-TAIJIc configuration yields the highest cross-correlation for isotropic SGWBs. These findings suggest that the relative geometry between space-based detectors is a key factor governing the scientific capabilities of detector networks. However, the impact of network geometry on the separation of tensor, vector, and scalar polarization components of SGWBs remains largely unexplored.

In this work, we investigate the capabilities of different LISA-TAIJI network configurations for probing polarized SGWBs. We focus on two proposed LISA-TAIJI networks with distinct relative constellation geometries and evaluate their performance in SGWB observations through cross-correlation measurements. First, we study the responses of the detector networks to tensor, vector, and scalar polarization modes and compare their sensitivities and signal-to-noise ratios (SNRs) for detecting and separating different polarization components. We then assess their ability to constrain polarized SGWB models described by power-law spectra within a Fisher-matrix framework. By comparing the results obtained from model-independent polarization-separation analyses and Fisher-matrix forecasts, we investigate how detector geometry influences different aspects of SGWB polarization measurements.

This paper is organized as follows. In Sec.~\ref{sec:networks}, we introduce the LISA-TAIJI network configurations and derive their responses to SGWBs. In Sec.~\ref{sec:sensitivity_snr}, we investigate the sensitivity, SNR, and polarization-separation capabilities of the two networks. In Sec.~\ref{sec:FIM_forecast}, we employ a Fisher-matrix framework to forecast constraints on SGWB components and compare the constraining power of the two network configurations. Finally, we summarize our conclusions and discuss future applications to space-based SGWB observations in Sec.~\ref{sec:conclusions}.

\section{Response of LISA-TAIJI networks to SGWB} \label{sec:networks}

\subsection{LISA-TAIJI network configurations}

The LISA mission consists of three spacecraft forming an approximately equilateral triangular constellation with an arm length of $2.5\times10^6~\mathrm{km}$ \cite{LISA:2017pwj,LISA:2024hlh}. The constellation follows a heliocentric orbit trailing the Earth by approximately $20^\circ$, while its formation plane is inclined by $60^\circ$ with respect to the ecliptic plane. TAIJI adopts a similar mission concept with a larger arm length of $3\times10^6~\mathrm{km}$ \cite{Hu:2017mde}.

\begin{figure}
\centering
\includegraphics[width=0.48\textwidth, trim=0cm 4cm 3cm 1cm, clip]{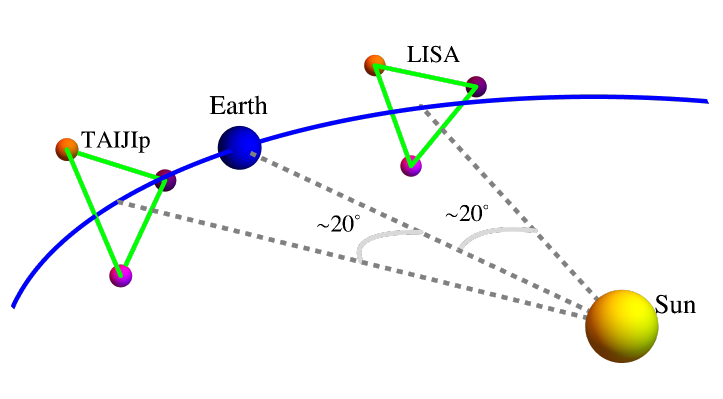}
\includegraphics[width=0.48\textwidth, trim=0cm 4cm 3cm 2cm, clip]{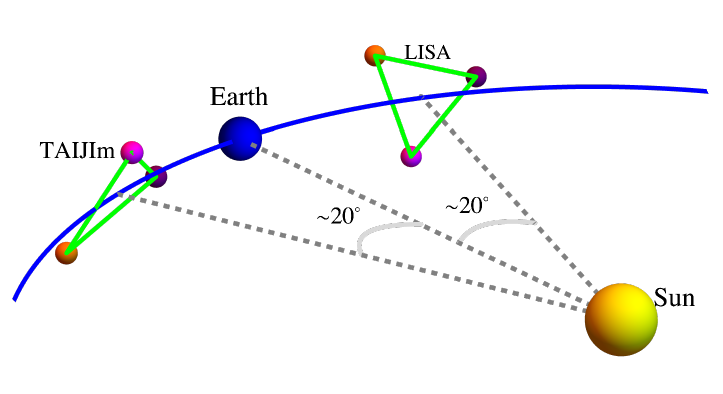}
\caption{\label{fig:LISA_TAIJI_orbit} Schematic illustration of the orbital configurations considered in this work. The upper panel shows the LISA--TAIJIp network, in which both constellation planes have positive inclinations relative to the ecliptic plane. The lower panel shows the LISA--TAIJIm network, where the TAIJI constellation adopts a negative inclination. As a result, the relative angle between the LISA and TAIJI constellation planes increases from approximately $34.5^\circ$ in the LISA--TAIJIp configuration to approximately $71^\circ$ in the LISA--TAIJIm configuration.
}
\end{figure}

In this work, we consider two representative LISA--TAIJI network configurations, namely LISA--TAIJIp and LISA--TAIJIm, whose orbital deployments are illustrated in Fig.~\ref{fig:LISA_TAIJI_orbit}. Both networks consist of the nominal LISA constellation and a TAIJI constellation leading the Earth by approximately $20^\circ$ along a heliocentric orbit. The two configurations differ only in the orientation of the TAIJI constellation plane. In the TAIJIp configuration, the TAIJI constellation is inclined by $+60^\circ$ relative to the ecliptic plane, whereas in the TAIJIm configuration the inclination is $-60^\circ$.

These different orientations result in substantially different relative geometries between the two detector constellations. The angle between the constellation planes of LISA and TAIJIp is approximately $34.5^\circ$, while the corresponding angle for LISA and TAIJIm is about $71^\circ$. In both configurations, the physical separation between the two constellations is approximately $D_{\rm sep}\sim10^8~\mathrm{km}$. Consequently, the two networks possess different antenna-pattern correlations, leading to different sensitivities for SGWB observations.

Previous investigations have shown that the large separation between the LISA and TAIJI constellations significantly improves source localization compared with a single detector \cite{Wang:2021uih}. Owing to the larger relative inclination between the detector planes, the LISA-TAIJIm network generally achieves more accurate parameter estimation than LISA-TAIJIp for massive black hole binary (MBBH) observations. On the other hand, the larger misalignment reduces the cross-correlation between the two detectors, particularly below $\sim 1$~mHz, leading to different sensitivities for SGWB observations \cite{Wang:2021njt}. The larger relative inclination, however, enhances the sensitivity of the network to parity-violating SGWB signals by providing a stronger distinction between the responses of the two detectors \cite{Chen:2024fto}.

The nearly colocated LISA-TAIJIc configuration provides the largest overlap reduction function (ORF) and therefore the highest sensitivity to an isotropic and unpolarized SGWB \cite{Wang:2021njt}. However, the strong correlation between the responses of the two colocated detectors substantially limits the ability of the network to distinguish different GW components. In addition, the short effective baseline between the two constellations leads to significantly poorer localization and parameter estimation capabilities for deterministic GW sources. Since the present work focuses on polarization separation and component characterization in SGWBs, we restrict our analysis to the LISA-TAIJIp and LISA-TAIJIm configurations, which provide more complementary detector responses and a broader range of network geometries.

\subsection{Time-delay interferometry} \label{secsub:TDI}

\begin{figure*}[htbp]
\includegraphics[width=0.9\textwidth]{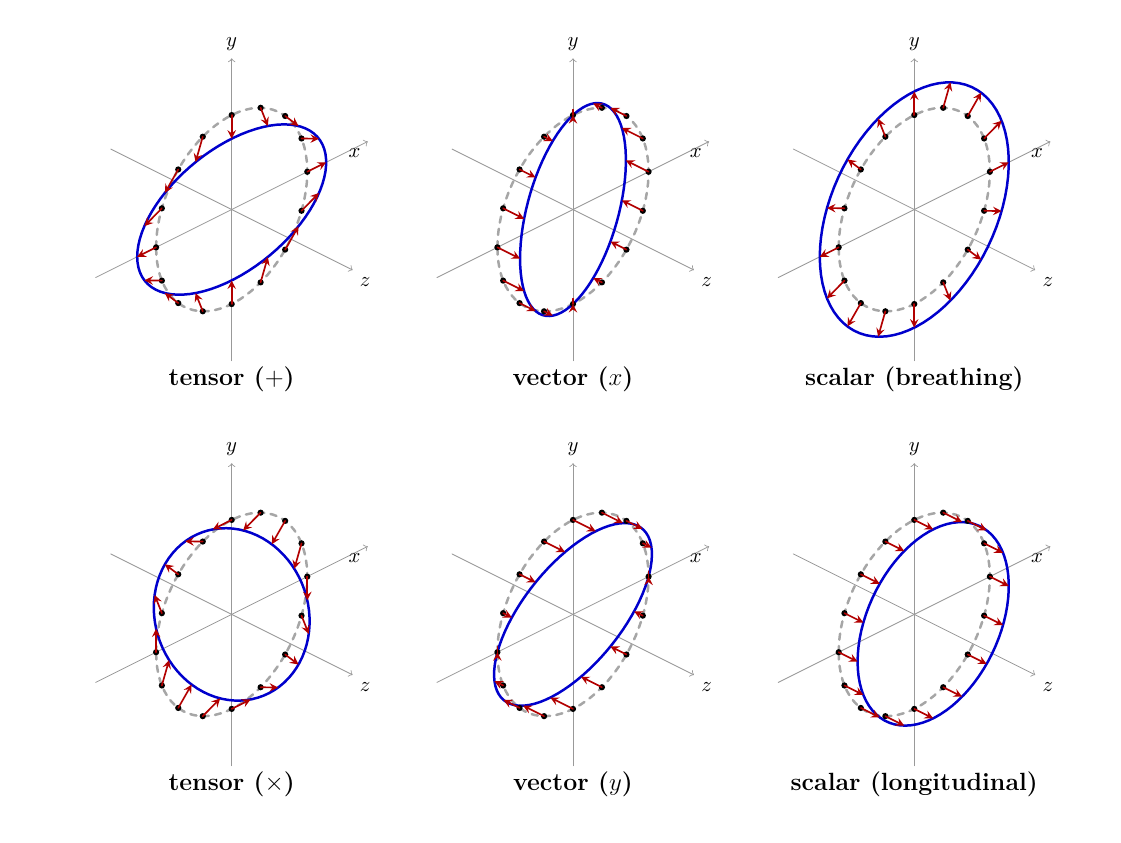}
\caption{Three-dimensional visualization of the tidal deformations produced by the six possible GW polarization states. The dashed gray circle denotes an initially circular ring of freely falling test masses in the transverse plane, while the solid blue curve shows the deformed configuration after the passage of a GW propagating along the $z$ axis. Red arrows indicate the displacement vectors of the individual test masses. Tensor plus ($+$), tensor cross ($\times$), and scalar breathing modes produce purely transverse deformations, whereas the vector $x$, vector $y$, and scalar longitudinal modes generate out-of-plane distortions through displacements parallel to the propagation direction. These characteristic deformation patterns underlie the distinct responses of GW detector networks to different polarization sectors.}
\label{fig:polarization_3d}
\end{figure*}

Different GW polarization states generate distinct tidal deformation patterns and therefore lead to different detector responses. In a general metric theory of gravity, GW may contain up to six independent polarization states, comprising the two tensor modes predicted by general relativity together with two vector and two scalar modes permitted by alternative theories of gravity. These polarization states produce qualitatively different tidal deformation patterns, which ultimately determine the response of a GW detector. Figure~\ref{fig:polarization_3d} illustrates the corresponding deformations of an initially circular ring of freely falling test masses for a wave propagating along the $z$ axis.

The tensor plus ($+$), tensor cross ($\times$), and scalar breathing modes generate purely transverse tidal fields, leaving the test masses confined to the $x$-$y$ plane. By contrast, the vector $x$, vector $y$, and scalar longitudinal modes induce displacements with nonvanishing components along the propagation direction, producing characteristic out-of-plane distortions. These distinct geometrical signatures lead to different antenna pattern functions and correlation for detector networks, providing the physical basis for separating tensor, vector, and scalar contributions in SGWB.

The response of a space-based GW detector is described in the solar system barycenter (SSB) reference frame \cite{Vallisneri:2012np,Katz:2022yqe,Baghi:2026fef}. The sky position of a GW source is specified by its ecliptic longitude $\lambda$ and ecliptic latitude $\beta$. The GW propagation direction is therefore given by
\begin{equation}
\mathbf{k} = - \mathbf{e}_r = ( -\cos \beta \cos \lambda, -\cos \beta \sin \lambda, -\sin \beta ).
\end{equation}
The corresponding orthonormal basis vectors along the longitude and latitude directions are
\begin{align}
\mathbf{u} = - \mathbf{e}_\phi = & ( \sin \lambda, - \cos \lambda, 0 ), \\
\mathbf{v} = - \mathbf{e}_\theta = & ( -\sin \beta \cos \lambda, -\sin \beta \sin \lambda, \cos \beta ).
\end{align}
Introducing the polarization angle $\psi$, the polarization basis vectors are obtained by rotating the $(\mathbf{u},\mathbf{v})$ basis according to
\begin{align}
 \mathbf{\epsilon}_1 = & \mathbf{u} \cos \psi + \mathbf{v} \sin \psi, \\
 \mathbf{\epsilon}_2 = & - \mathbf{u} \sin \psi + \mathbf{v} \cos \psi.
\end{align}
The polarization tensors for the six possible GW polarization modes are then given by
\begin{align}
 \mathbf{e}_+ = & \mathbf{\epsilon}_1 \otimes \mathbf{\epsilon}_1 - \mathbf{\epsilon}_2 \otimes \mathbf{\epsilon}_2, \\
 \mathbf{e}_\times = & \mathbf{\epsilon}_1 \otimes \mathbf{\epsilon}_2 + \mathbf{\epsilon}_2 \otimes \mathbf{\epsilon}_1, \\
 \mathbf{e}_x = & \mathbf{\epsilon}_1 \otimes \mathbf{e}_r + \mathbf{e}_r \otimes \mathbf{\epsilon}_1, \\
 \mathbf{e}_y = & \mathbf{\epsilon}_2 \otimes \mathbf{e}_r + \mathbf{e}_r \otimes \mathbf{\epsilon}_2, \\
 \mathbf{e}_b = & \mathbf{\epsilon}_1 \otimes \mathbf{\epsilon}_1 + \mathbf{\epsilon}_2 \otimes \mathbf{\epsilon}_2, \\
 \mathbf{e}_l = & \sqrt{2} \mathbf{e}_r \otimes \mathbf{e}_r.
\end{align}
For a laser link connecting a transmitting spacecraft at $\mathbf{r}_s$ and a receiving spacecraft at $\mathbf{r}_r$, the projected GW strain associated with polarization mode $p$ can be written as
\begin{equation}
\begin{aligned}
    h_{\mathrm{SSB}, p} (t, \mathbf{r} ) = &  \mathbf{n}_{sr} \cdot \mathbf{e}_p \cdot \mathbf{n}^T_{sr} \  H_{p}(t, \mathbf{r} ),
\end{aligned}
\end{equation}
where $\mathbf n_{sr}$ denotes the unit vector from the transmitting spacecraft to the receiving spacecraft, and $H_p$ represents the metric perturbation associated with polarization mode $p$.
The corresponding fractional frequency fluctuation induced by the GW is
\begin{equation}
    y_{sr, p}(f, \Omega) = \frac{ \mathbf{n}_{sr} \cdot \mathbf{e}_p \cdot \mathbf{n}^T_{sr} }{2 \left[ 1 - \mathbf{k} \cdot \mathbf{n}_{sr} \right]} \left[ e^{ 2 \pi i f ( \mathbf{k} \cdot \mathbf{r}_s - \tau_s ) } - e^{ 2 \pi i f ( \mathbf{k} \cdot \mathbf{r}_r - \tau_r ) } \right],
\end{equation}
where $\Omega = (\lambda,\beta,\psi)$ collectively denotes the angular parameters specifying the source direction and polarization state.

Time-delay interferometry (TDI) is adopted in both LISA and TAIJI to suppress laser frequency noise to a level below the secondary noise sources and thereby achieve the designed sensitivity. TDI constructs virtual equal-arm interferometers by combining appropriately time-shifted inter-spacecraft phase measurements, effectively cancelling the otherwise dominant laser frequency fluctuations.
A wide array of second-generation TDI combinations has been introduced in existing literature, with each architectural design offering distinct operational characteristics. These formulations typically diverge in their overall delay durations, the analytical behavior of their transfer functions, and the specific locations of their null frequencies \cite{Wang:2025voa}.

In this work, we employ the PD4L TDI configuration introduced in \cite{Wang:2011tlj,Wang:2025voa}. PD4L is a second-generation TDI observable that can be constructed from combinations of the first-generation Monitor and Beacon variables. It consists of 16 inter-spacecraft links and satisfies the requirements for laser-noise cancellation in a constellation with time-varying arm lengths. The optical paths of the three ordinary PD4L observables are
\begin{align}
\text{PD4L1} &: \overrightarrow{1 2 3 2} \ \overleftarrow{2 1 2} \ \overrightarrow{2 3 2 1} \ \overleftarrow{1 3 2 3} \ \overrightarrow{3 1 3} \ \overleftarrow{3 2 3 1}\,, \label{eq:PD4L-1} \\
\text{PD4L2} &: \overrightarrow{2 3 1 3} \ \overleftarrow{3 2 3} \ \overrightarrow{3 1 3 2} \ \overleftarrow{2 1 3 1} \ \overrightarrow{1 2 1} \ \overleftarrow{1 3 1 2}\,, \label{eq:PD4L-2} \\
\text{PD4L3} &: \overrightarrow{3 1 2 1} \ \overleftarrow{1 3 1} \ \overrightarrow{1 2 1 3} \ \overleftarrow{3 2 1 2} \ \overrightarrow{2 3 2} \ \overleftarrow{2 1 2 3}\, \label{eq:PD4L-3} .
\end{align}
The path notation follows the convention of \citet{Vallisneri:2005ji}. Integer labels denote spacecraft indices, while the arrows indicate the temporal ordering of the light propagation sequence. A right-pointing arrow corresponds to a forward-time evolution, whereas a left-pointing arrow denotes a sequence evaluated backward in time.
The response of a TDI observable is obtained by combining the responses of the constituent inter-spacecraft links \cite{1975GReGr...6..439E,1987GReGr..19.1101W,Vallisneri:2007xa,Vallisneri:2012np,Tinto:2010hz}, the response function for a specific polarization mode $p$ is given by
\begin{equation}
\begin{aligned}
 F_p (f, \Omega) = & \sum_{sr, k+}  {y}_{sr, p} \ e^{2 \pi i f \tau_{k+}} - \sum_{sr, k-}  {y}_{sr, p} \ e^{2 \pi i f \tau_{k-} },
 \end{aligned}
\end{equation}
where the first summation corresponds to one synthesized optical path (indicated by the right-pointing arrows in Eqs.~\eqref{eq:PD4L-1}--\eqref{eq:PD4L-3}), and the second summation corresponds to the counter-propagating path (indicated by the left-pointing arrows). The quantities $\tau_{k+}$ and $\tau_{k-}$ denote the cumulative time delays associated with the $k$th link in the two synthesized paths, respectively, and $p \in \{+, \times, x, y, b, l\}$ labels the GW polarization mode.

Compared with the conventional second-generation Michelson observables, PD4L has a shorter effective time span. While the Michelson combinations involve an overall delay duration of $8L$, the corresponding duration for PD4L is $4L$, where $L$ denotes the average one-way light travel time. Previous studies have shown that PD4L has a lower density of transfer-function nulls and mitigates edge effects associated with finite-duration data segments \cite{Wang:2025mee,Wang:2025voa}. Since the present work focuses on SGWB observations with numerically generated TDI responses, PD4L is adopted throughout the analysis.

For a LISA-like detector with six inter-spacecraft laser links, three optimal TDI channels (A, E, T) could be constructed from three second-generation TDI PD4L observables (PD4L1, PD4L2, PD4L3) \cite{Prince:2002hp,Vallisneri:2007xa}:
\begin{equation} \label{eq:optimalTDI}
\begin{bmatrix}
\mathrm{A}  \\ \mathrm{E}  \\ \mathrm{T}
\end{bmatrix}
 =
\begin{bmatrix}
-\frac{1}{\sqrt{2}} & 0 & \frac{1}{\sqrt{2}} \\
\frac{1}{\sqrt{6}} & -\frac{2}{\sqrt{6}} & \frac{1}{\sqrt{6}} \\
\frac{1}{\sqrt{3}} & \frac{1}{\sqrt{3}} & \frac{1}{\sqrt{3}}
\end{bmatrix}
\begin{bmatrix}
\mathrm{PD4L1} \\ \mathrm{PD4L2} \\ \mathrm{PD4L3}
\end{bmatrix}.
\end{equation}
where the A and E channels provide the primary GW sensitivity, while the T channel behaves as an approximate null channel at low frequencies and is dominated by instrumental noise.
The ORF quantifies the correlated response of two TDI channels to an isotropic SGWB and plays a central role in cross-correlation analyses. For tensor, vector, and scalar polarizations, the ORFs are defined as \cite{Nishizawa:2009bf,Romano:2016dpx}
\begin{align}
\Gamma^{\rm (T)}_{ij} = & \frac{1}{4 \pi} \int d^2 \Omega \left[ F_{+,i}(f, \mathbf{n}) F^{\ast}_{+,j} (f, \mathbf{n}) + F_{\times, i} (f, \mathbf{n}) F^{\ast}_{\times , j} (f, \mathbf{n}) \right] \label{eq:ORF_T} \\
 \Gamma^{\rm (V)}_{ij} = & \frac{1}{4 \pi} \int d^2 \Omega \left[ F_{x,i}(f, \mathbf{n}) F^{\ast}_{x, j} (f, \mathbf{n}) + F_{y , i} (f, \mathbf{n}) F^{\ast}_{y, j} (f, \mathbf{n}) \right] \label{eq:ORF_V} \\
\Gamma^{\rm (S)}_{ij} = & \frac{1}{4 \pi} \int d^2 \Omega \left[ F_{b,i}(f, \mathbf{n}) F^{\ast}_{b, j} (f, \mathbf{n}) + F_{l , i} (f, \mathbf{n}) F^{ \ast}_{l, j} (f, \mathbf{n}) \right] \label{eq:ORF_S}
\end{align}

After laser frequency noise is adequately suppressed by TDI, the dominant instrumental noise sources are the test-mass acceleration noise and the optical metrology system (OMS) noise. These noise contributions are included in the sensitivity calculations presented below \cite{LISA:2017pwj,LISA:2024hlh,Luo:2020}.
The acceleration-noise requirements of LISA and TAIJI are assumed to be identical \cite{LISA:2017pwj,LISA:2024hlh,Luo:2020},
\begin{equation}
 S_{\rm acc} = \frac{ (3 \ \rm fm/s^2)^2 }{ \rm Hz } \left[ 1 + \left( \frac{0.4 \ {\rm mHz}}{f} \right)^2 \right]  
 \left[ 1 + \left(\frac{f}{8 \ {\rm mHz}} \right)^4 \right] .
\end{equation}
The OMS noise requirements differ slightly between the two missions,
\begin{align}
 S_{\rm oms, LISA} & = \frac{ (15 \ \rm pm)^2}{\rm Hz} \left[ 1 + \left(\frac{2 \ {\rm mHz}}{f} \right)^4 \right],  \\
S_{\rm oms, TAIJI} & = \frac{ (8 \ \rm pm)^2}{\rm Hz} \left[ 1 + \left(\frac{2 \ {\rm mHz}}{f} \right)^4 \right].
 \end{align}
The noise power spectral densities (PSDs) of the TDI observables are evaluated numerically \cite{Wang:2020fwa,Wang:2ndTDI}.

\section{Polarization separation with LISA-TAIJI networks} \label{sec:sensitivity_snr}

\subsection{Cross-correlation formalism}

The separation of different SGWB polarization components relies on the cross-correlation of data streams from spatially separated detectors. In a detector network, the correlated SGWB signal can be distinguished from the uncorrelated instrumental noise through the ORFs, which encode the relative geometry and polarization responses of the detectors \cite{Allen:1996vm,Allen:1997ad,Romano:2016dpx}. In this work, we adopt the standard cross-correlation formalism and evaluate the performance of the LISA-TAIJI networks in detecting and separating tensor, vector, and scalar polarization components.

The fundamental observable is the cross-correlation estimator constructed from two TDI channels,
\begin{equation}
    C_{ij} (f) = \frac{2}{T_{\rm obs}} \tilde{d}_{i}(f) \tilde{d}_{j}^*(f) \,,
\end{equation}
where $T_{\rm obs}$ denotes the observation duration and $\tilde d_i(f)=\tilde h_i(f)+\tilde n_i(f)$ is the Fourier-domain data stream of TDI channel $i \in {\mathrm{A,E,T}}$, consisting of both the SGWB signal and instrumental noise contributions.
Assuming that the instrumental noises in different detectors are statistically independent, the ensemble average of the cross-correlation estimator is determined solely by the SGWB signal,
\begin{equation} \label{eq:C_average}
\begin{aligned}
 \left\langle C_{ij} \right\rangle & = \left\langle s_i, s_j \right\rangle + \left\langle s_i , n_j \right\rangle  + \left\langle n_i , s_j \right\rangle + \left\langle n_i , n_j \right\rangle \\
 & = \left\langle s_i , s_j \right\rangle \\
 & = \sum_{P}  \Gamma_{ij}^{(P)} (f) S_h^{(P)} (f).
 \end{aligned}
\end{equation}
where $P\in{\mathrm{T,V,S}}$ labels the tensor, vector, and scalar polarization sectors, respectively. Throughout this work, parity-violating components are not considered. The quantities $S_h^{(P)}(f)$ and $\Gamma_{ij}^{(P)}(f)$ denote the strain PSD and ORF of polarization mode $P$.
The expected squared SNR, denoted as $\rho^2$, is conventionally defined as the squared expectation of the cross-correlation divided by its variance \cite[and references therein]{Allen:1996vm,Allen:1997ad,Cornish:2001qi}:
\begin{equation}
\rho^2  = \frac{  \left\langle C \right\rangle^2 }{ \left\langle \mathcal{N}^2 \right\rangle} = \frac{  \left\langle C \right\rangle^2 }{ \left\langle C^2 \right\rangle - \left\langle C \right\rangle^2 } \,.
\end{equation}
Based on the optimal filter, the maximized SNR for observing the polarization mode $P$ through the correlation of two TDI channels becomes
\begin{equation}  \label{eq:rho2_origin}
 \rho_{(P)}^2 =  T_\mathrm{obs} \int^{+\infty}_{-\infty} \mathrm{d} f \frac{ \left[ \Gamma_{ij}^{(P)} (f) S_h^{(P)} (f) \right]^2 }{ M(f) },
\end{equation}
where the effective variance density is
\begin{widetext}
\begin{equation} \label{eq:M_factor}
M =  \left[ N_i (f) + \sum_P S^{(P)}_{h} (f) \Gamma^{(P)}_{ii} (f) \right] \left[ N_j (f) + \sum_P S^{(P)}_{h} (f) \Gamma^{(P)}_{jj} (f) \right]  +  \left[ \sum_{P} \Gamma_{ij}^{(P)} (f) S_h^{(P)} (f) \right]^2.
\end{equation}
\end{widetext}
Here $N_i$ denotes the noise PSD of TDI observable $i$. The first term describes the combined contribution of instrumental noise and SGWB self-correlations in the two channels, while the second term accounts for the variance associated with the correlated SGWB signal itself.

For the weak-signal regime relevant to most SGWB searches, $N_i(f)\gg S_h^{(P)}(f)$, the instrumental noise dominates over the SGWB contribution and the variance density reduces to $M(f)\simeq N_i(f)N_j(f)$. The network SNR obtained by combining all independent TDI-channel pairs can be approximated as
\begin{equation} \label{eq:rho2}
\begin{aligned}
 \rho_{(P)}^2 & \simeq \sum_{i,j \in \{\mathrm{A, E, T}\} }  T_\mathrm{obs} \int^{+\infty}_{-\infty} \mathrm{d} f \frac{ \left[ \Gamma_{ij}^{(P)} (f)  S_h^{(P)} (f) \right]^2  }{N_{i} (f) N_{j} (f) } \\
 & =  \sum_{i,j \in \{\mathrm{A, E, T}\} } 2 T_\mathrm{obs} \int^{+ \infty }_{0} \mathrm{d} f  \frac{ \left[ \Gamma_{ij}^{(P)} (f)  S_h^{(P)} (f) \right]^2 }{ N_{i} (f)  N_{j} (f) }.
\end{aligned}
\end{equation}
Throughout this section, we adopt an effective observation time of $T_{\rm obs}=3~{\rm yr}$, corresponding to the nominal four-year mission duration and an assumed duty cycle of 75\% \cite{LISA:2017pwj,Caprini:2019pxz}.

\subsection{Sensitivity for polarization component separation} \label{sec:separation_sensitivity}

To assess the capability of a LISA-TAIJI network to separate different polarization components of a SGWB, we adopt the component-separation formalism developed in Refs.~\cite{Nishizawa:2009bf,Romano:2016dpx}. The analysis is based on the set of cross-correlation measurements obtained from all independent detector-channel pairs. For a given signal model, the likelihood of the measured cross-correlation data can be written as
\begin{equation}
    p(\hat{\bm{C}} | \bm{\mathcal{A}}) \propto \exp\left[ -\frac{1}{2} (\hat{\bm{C}} - \bm{\mathcal{M}}\bm{\mathcal{A}})^\dagger \bm{\mathcal{N}}^{-1} (\hat{\bm{C}} - \bm{\mathcal{M}} \bm{\mathcal{A}}) \right] \,,
\end{equation}
where the response matrix of the detector network and the signal vector are defined as
\begin{equation} \label{eq:M_mat_A_vec}
    \bm{\mathcal{M}} = 
    \begin{bmatrix}
        \Gamma_{12}^{(T)} & \Gamma_{12}^{(V)} & \Gamma_{12}^{(S)} \\
        \Gamma_{13}^{(T)} & \Gamma_{13}^{(V)} & \Gamma_{13}^{(S)} \\
        \vdots & \vdots & \vdots \\
        \Gamma_{m-1 m}^{(T)} & \Gamma_{m-1 m}^{(V)} & \Gamma_{m-1 m}^{(S)}
    \end{bmatrix}, \, \quad
    \bm{\mathcal{A}} = 
    \begin{bmatrix}
        S_h^{(T)} \\
        S_h^{(V)} \\
        S_h^{(S)}
    \end{bmatrix} \,.
\end{equation}
Maximizing the likelihood with respect to the signal vector $\bm{\mathcal A}$ yields the maximum likelihood estimator, $\hat{\bm{\mathcal{A}}} = \bm{F}^{-1} \bm{X}$, where $\bm{X} = \bm{\mathcal{M}}^\dagger \bm{\mathcal{N}}^{-1} \hat{\bm{C}}$ and $\bm{F}$ is the matrix defined as $\bm{F} = \bm{\mathcal{M}}^\dagger \bm{\mathcal{N}}^{-1} \bm{\mathcal{M}}$. 
For a multi-baseline detector network, the matrix $\bm F$ at a given frequency can be written as
\begin{equation} \label{eq:F_matrix_three_SGWB}
    \bm{F} = 
    \begin{bmatrix}
        \sum_{ij} \frac{(\Gamma_{ij}^{(T)})^2}{N_i N_j} & \sum_{ij} \frac{\Gamma_{ij}^{(T)}\Gamma_{ij}^{(V)}}{N_i N_j} & \sum_{ij} \frac{\Gamma_{ij}^{(T)}\Gamma_{ij}^{(S)}}{N_i N_j} \\[1.5em]
        \sum_{ij} \frac{\Gamma_{ij}^{(V)}\Gamma_{ij}^{(T)}}{N_i N_j} & \sum_{ij} \frac{(\Gamma_{ij}^{(V)})^2}{N_i N_j} & \sum_{ij} \frac{\Gamma_{ij}^{(V)}\Gamma_{ij}^{(S)}}{N_i N_j} \\[1.5em]
        \sum_{ij} \frac{\Gamma_{ij}^{(S)}\Gamma_{ij}^{(T)}}{N_i N_j} & \sum_{ij} \frac{\Gamma_{ij}^{(S)}\Gamma_{ij}^{(V)}}{N_i N_j} & \sum_{ij} \frac{(\Gamma_{ij}^{(S)})^2}{N_i N_j}
    \end{bmatrix} \,.
\end{equation}
For a LISA-TAIJI network, nine independent cross-correlation baselines can be constructed from the three optimal TDI channels $\{\mathrm{A,E,T}\}$ of each detector.

The covariance matrix of the recovered polarization components is given by, $\langle \hat{\bm{\mathcal{A}}} \hat{\bm{\mathcal{A}}}^\dagger \rangle - \langle \hat{\bm{\mathcal{A}}} \rangle \langle \hat{\bm{\mathcal{A}}}^\dagger \rangle \approx \bm{F}^{-1}$.
Provided that $\det(\bm F)\neq 0$, the variances of the recovered tensor, vector, and scalar components are given by the diagonal elements of $\bm F^{-1}$,
\begin{align}
    \sigma_{\hat{T}}^2 &= (\bm{F}^{-1})_{11}, 
    \label{eq:var_T}  \\
    \sigma_{\hat{V}}^2 &= (\bm{F}^{-1})_{22}, 
    \label{eq:var_V} \\
    \sigma_{\hat{S}}^2 &= (\bm{F}^{-1})_{33}. 
    \label{eq:var_S}
\end{align}
These variances can be translated into effective sensitivities for the corresponding polarization components,
\begin{equation}
\Omega^{(P)}_\mathrm{eff} = \frac{4 \pi^2 f^3}{3 H^2_0} \sigma_{P}
\end{equation}
where $H_0 \simeq 2.185 \times 10^{-18}$ Hz is the Hubble constant \cite{Planck:2018vyg}. 
When only two polarization sectors are considered, for instance tensor-scalar (TS) or tensor-vector (TV) scenario, Eqs. \eqref{eq:M_mat_A_vec}-\eqref{eq:var_S} could be easily modified by reducing the dimension from three to two. 
For a pure polarization background, the formalism reduces to the standard SGWB sensitivity calculation.

To provide a model-independent characterization of the network sensitivity, we additionally compute the power-law integrated sensitivity (PLS)~\cite{Thrane:2013oya}. Assuming a power-law SGWB spectrum, $\Omega_h(f)=A_i \left(\frac{f}{f_{\rm ref}}\right)^{\alpha_i}$, the PLS corresponds to the envelope of all detectable power-law spectra for a given observation time and detection threshold,
\begin{align}
\Omega_{i}  = & \frac{\rho_\mathrm{th}}{\sqrt{2 T_\mathrm{obs}}} \frac{4 \pi^2 }{3 H^2_0}   \left[  \int^{f_\mathrm{max}}_0 \mathrm{d} f \frac{ (f/f_\mathrm{ref})^{2 \alpha_i} }{ f^6 \sigma^2_P } \right]^{-1/2}, \\
 \Omega_\mathrm{PLS} & = \max _{\alpha_i} \left[ \Omega_i \left( \frac{f}{f_\mathrm{ref}} \right)^{\alpha_i} \right], \quad \mathrm{for} \ \alpha_i \in [-8,  8].
\end{align}
The SNR threshold is set to be $\rho_\mathrm{th} =  10$ in this calculation.

\begin{figure*}[htbp]
\includegraphics[width=0.48\textwidth]{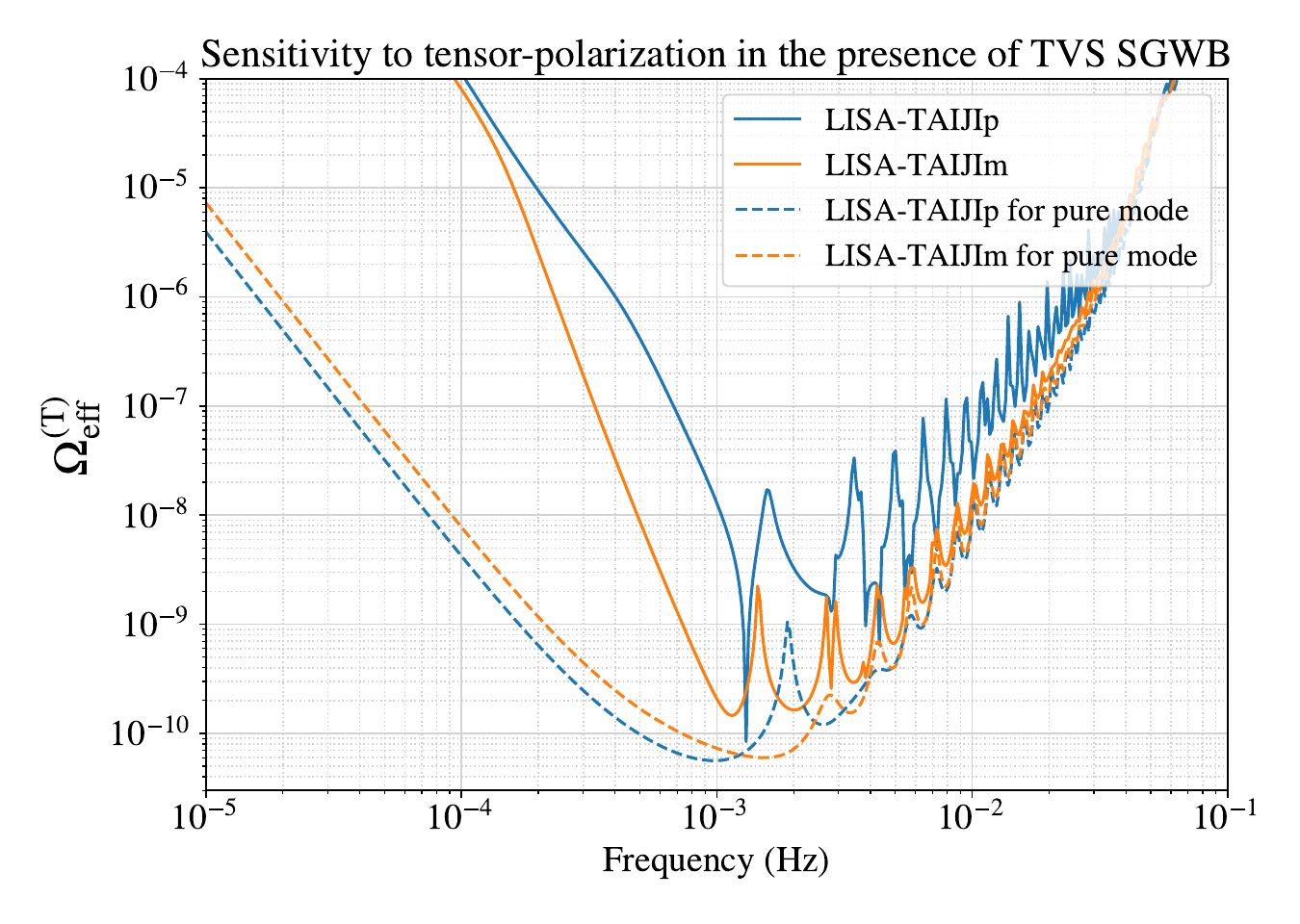}
\includegraphics[width=0.48\textwidth]{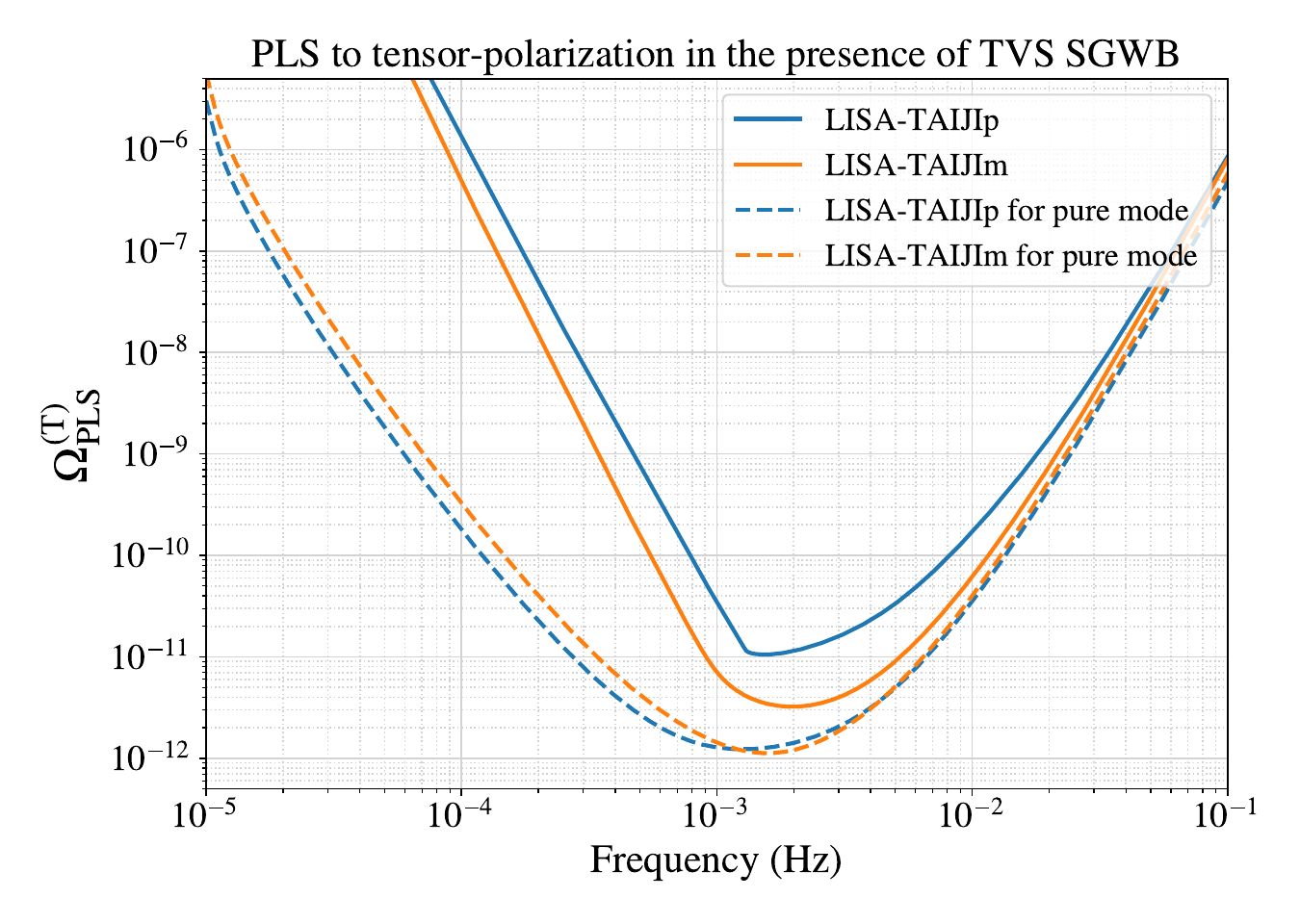} 
\includegraphics[width=0.48\textwidth]{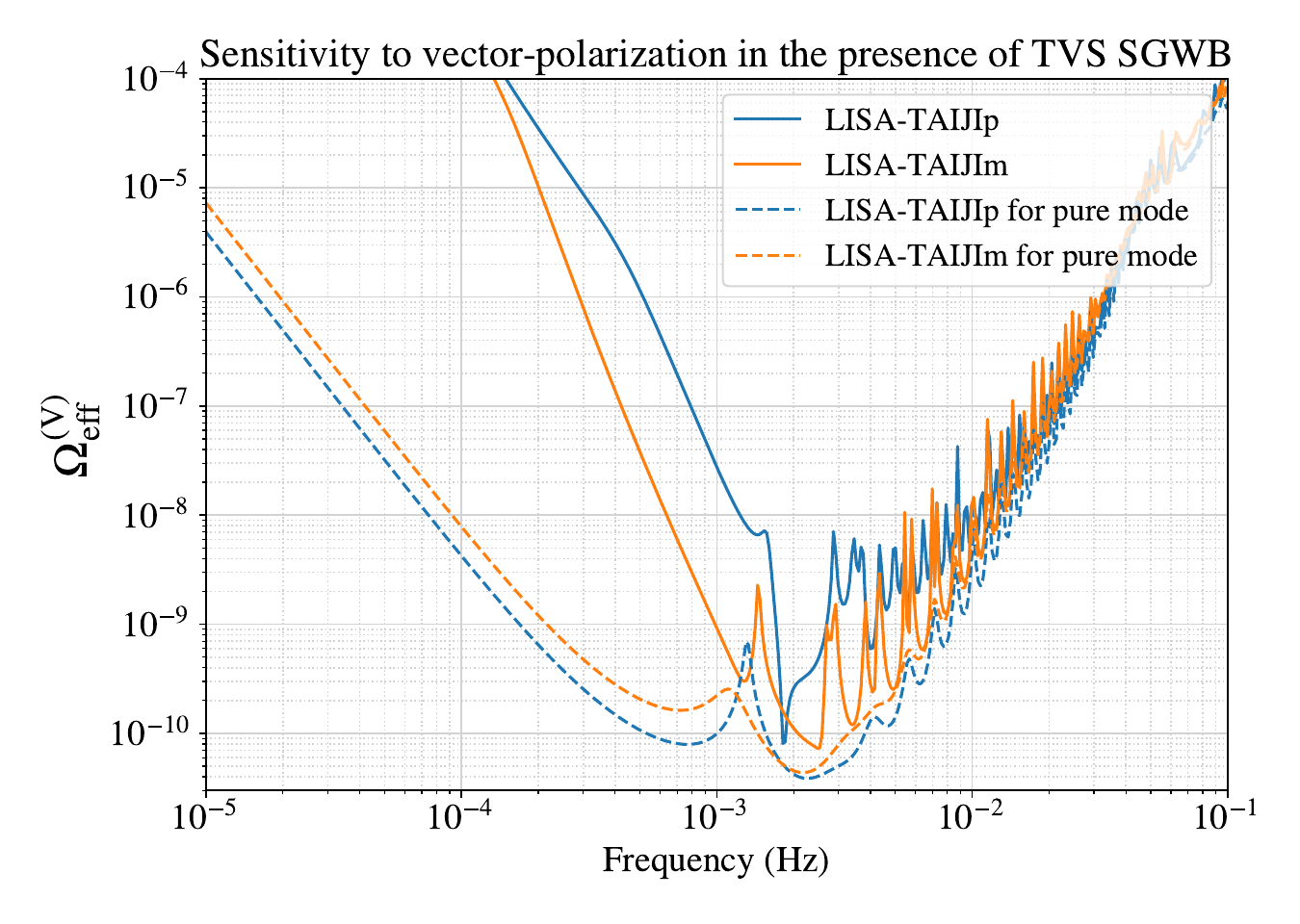}
\includegraphics[width=0.48\textwidth]{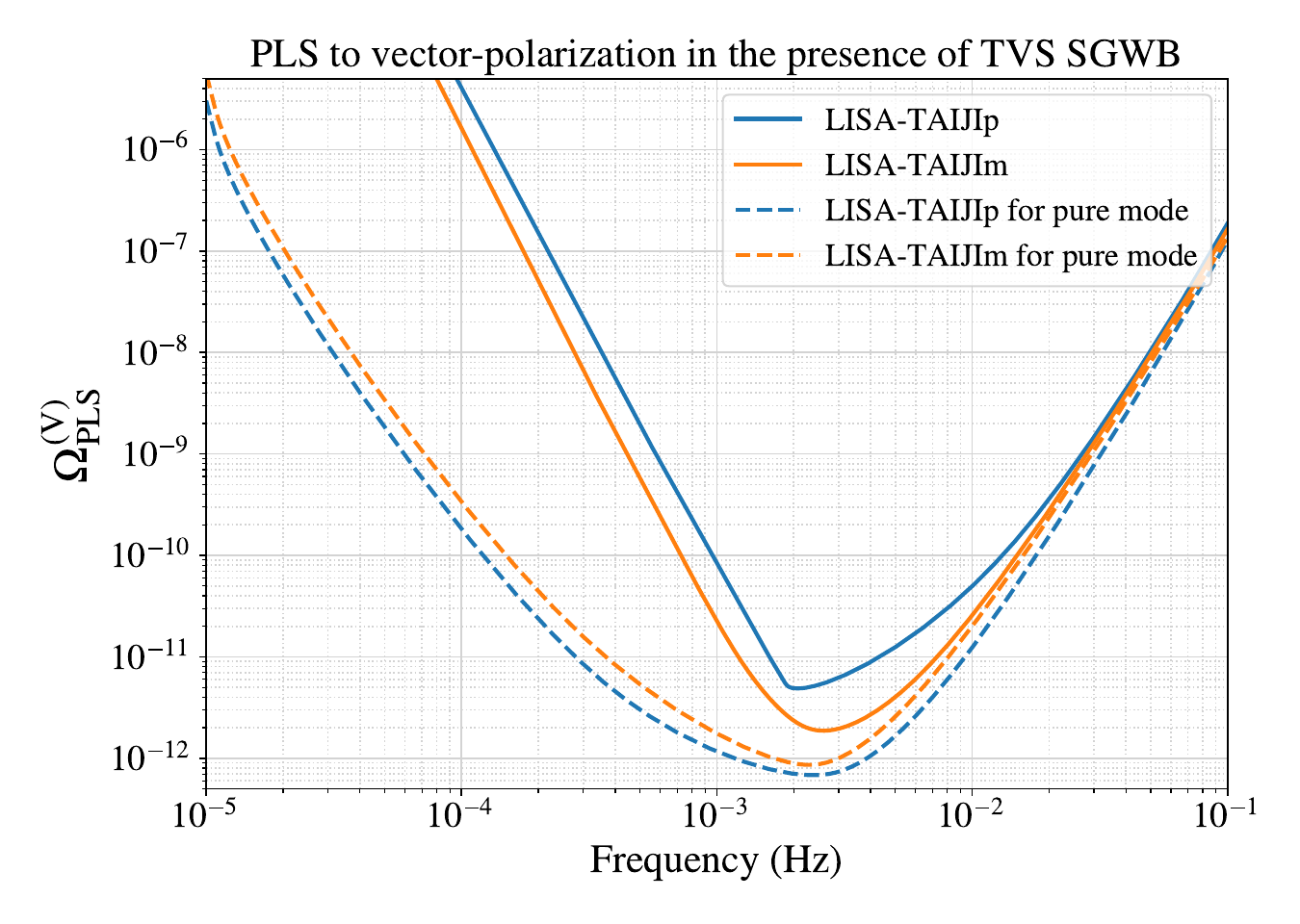}
\includegraphics[width=0.48\textwidth]{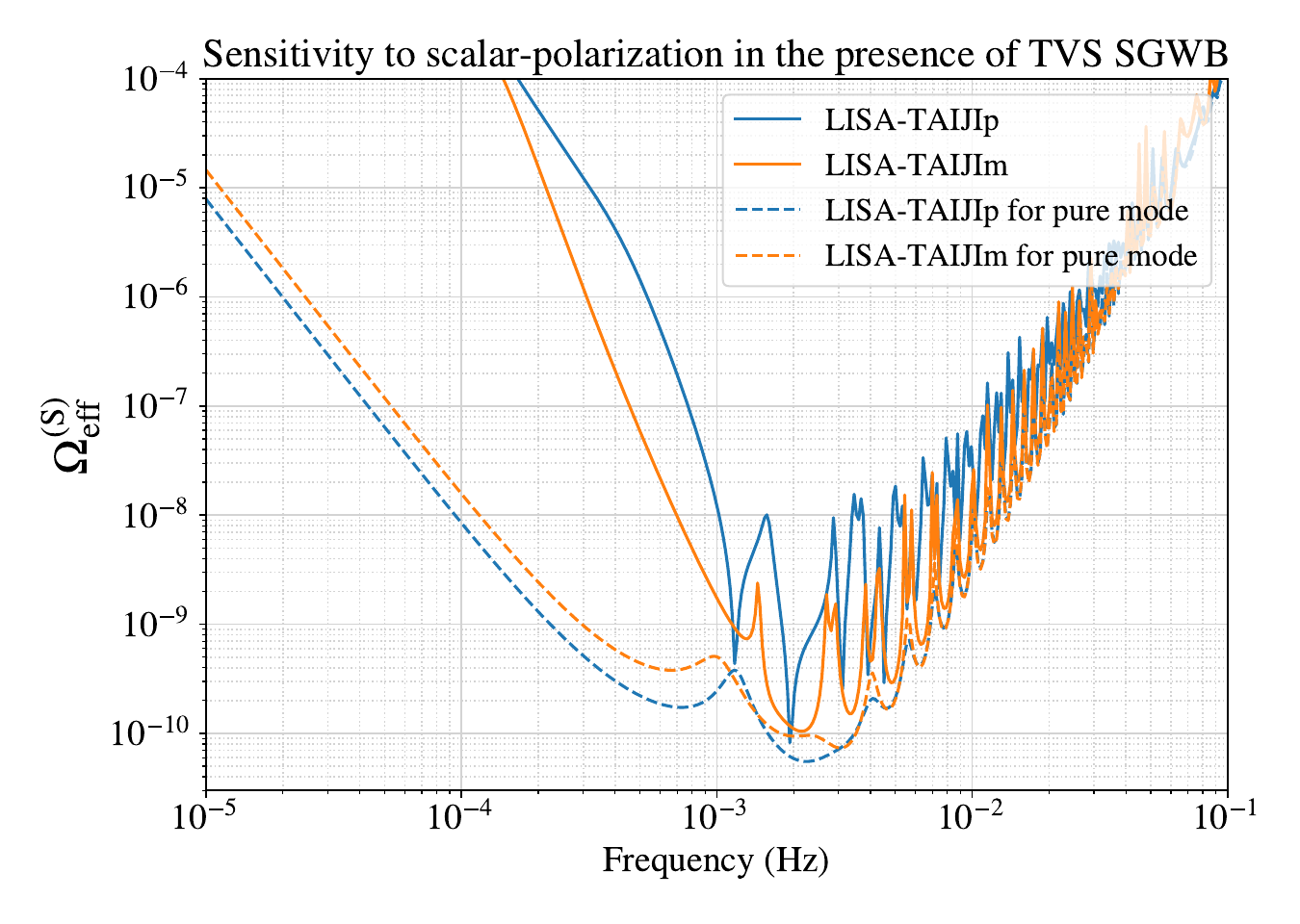} 
\includegraphics[width=0.48\textwidth]{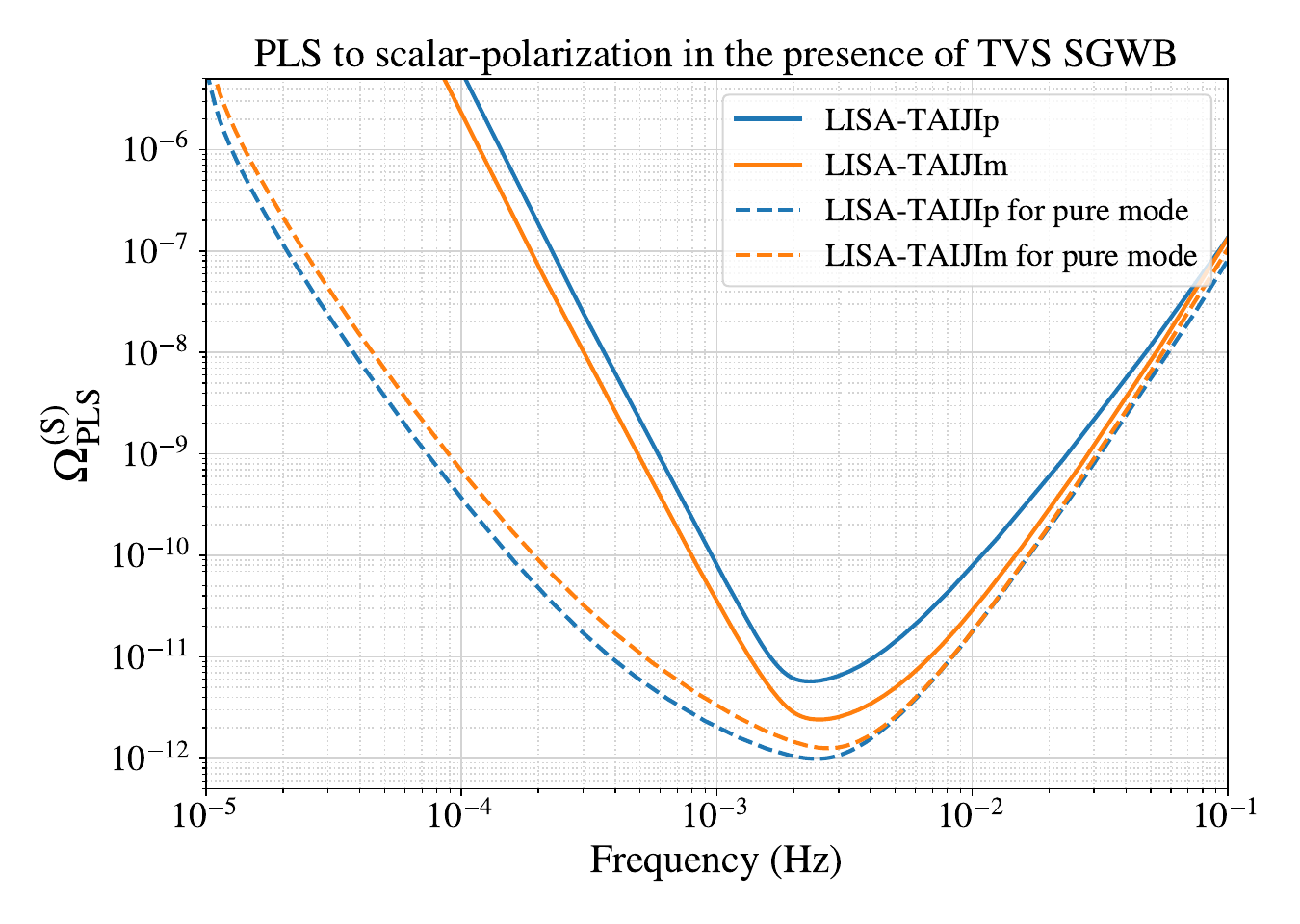} 
\caption{\label{fig:sensitivity_LISA_TAIJI_TVS} Sensitivities of the LISA-TAIJI detector networks to the tensor (top row), vector (middle row), and scalar (bottom row) polarization components of a SGWB containing simultaneous tensor, vector, and scalar modes. The left panels show the effective fractional energy-density sensitivities, $\Omega_{\rm eff}^{(P)}(f)$, while the right panels display the corresponding PLSs, $\Omega_{\rm PLS}^{(P)}$, assuming a detection threshold of $\rho_{\rm th}=10$ and an observation time of $T_{\rm obs}=3$ yr. Blue and orange curves correspond to the LISA-TAIJIp and LISA-TAIJIm configurations, respectively. Dashed curves denote the idealized pure-polarization sensitivities obtained when only the target polarization mode is present, whereas solid curves include the simultaneous presence of tensor, vector, and scalar components and therefore account for polarization degeneracies in the network response. The larger separation between the dashed and solid curves for the LISA-TAIJIp configuration indicates stronger polarization degeneracies, whereas the LISA-TAIJIm network retains a greater fraction of its pure-mode sensitivity when all three polarization sectors are simultaneously present.
}
\end{figure*}

\begin{figure*}[htbp]
\includegraphics[width=0.48\textwidth]{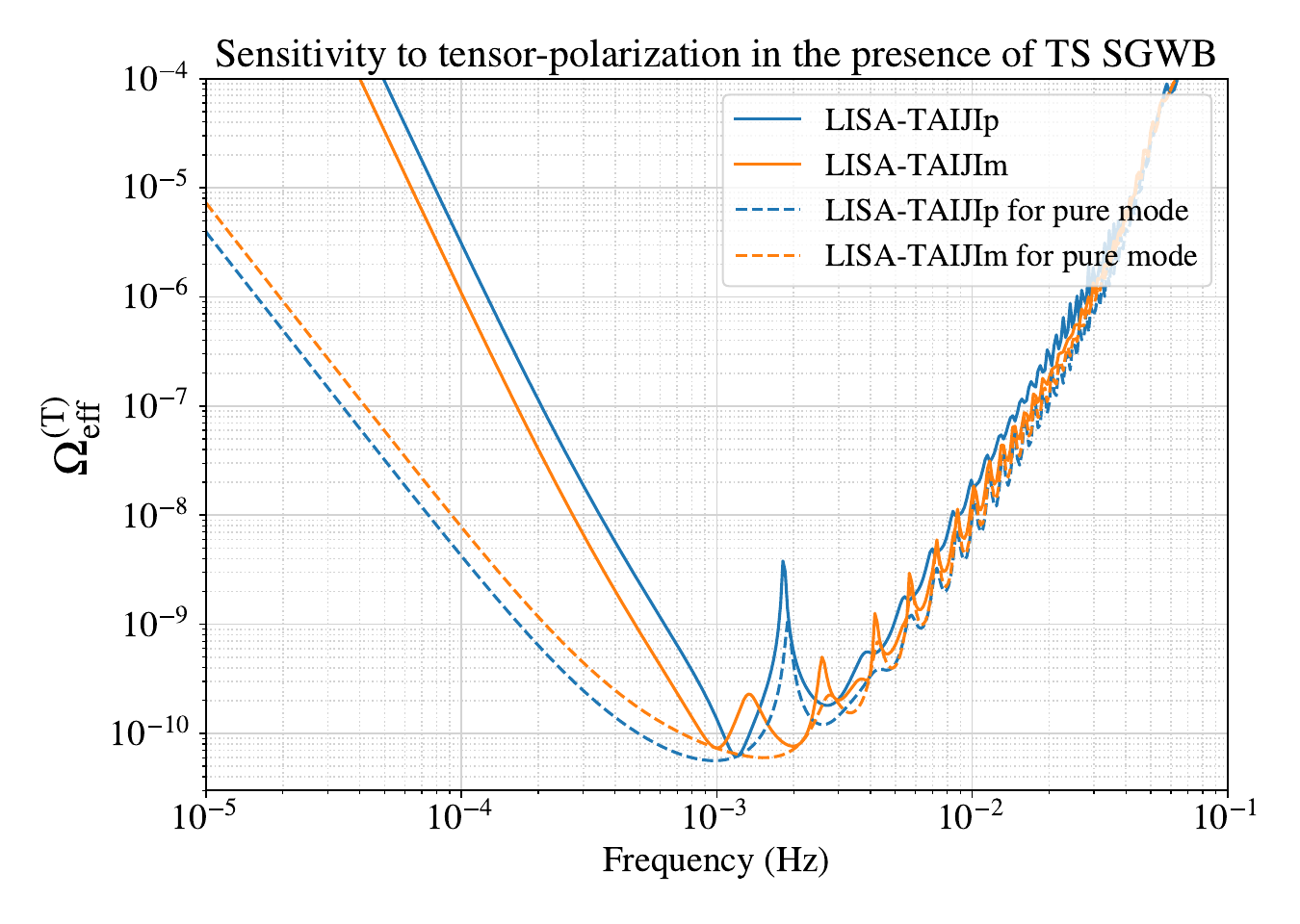} 
\includegraphics[width=0.48\textwidth]{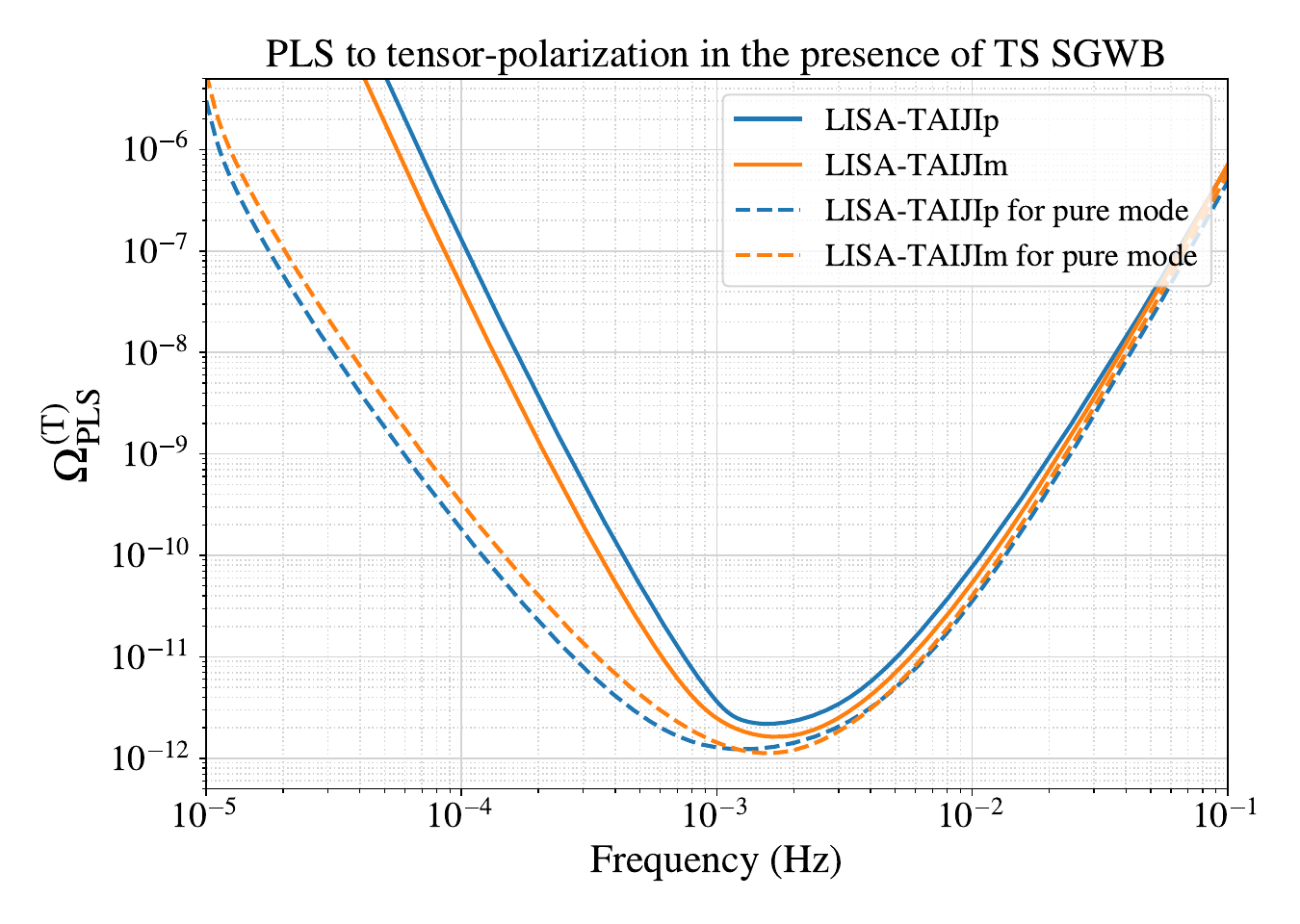} 
\includegraphics[width=0.48\textwidth]{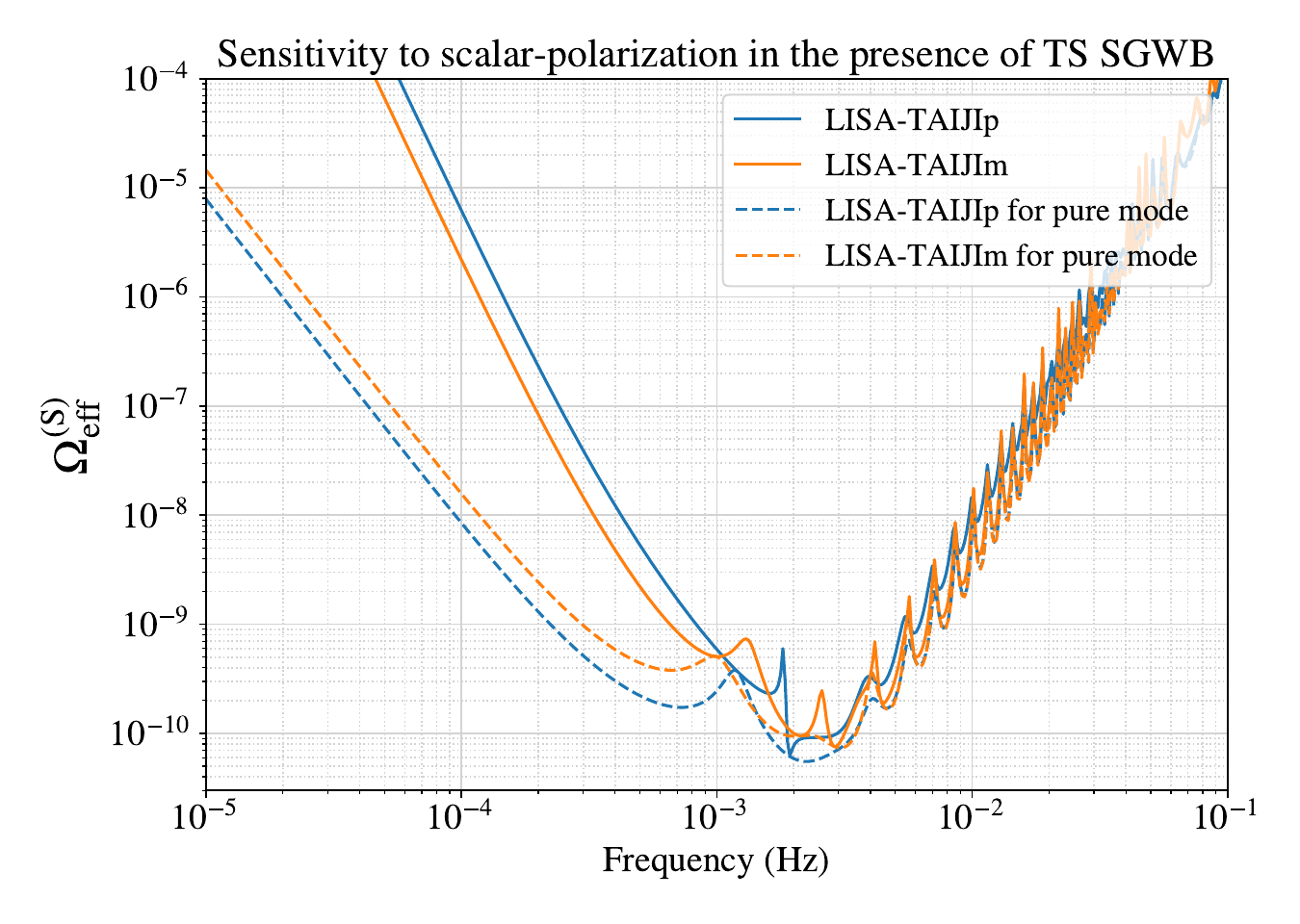}
\includegraphics[width=0.48\textwidth]{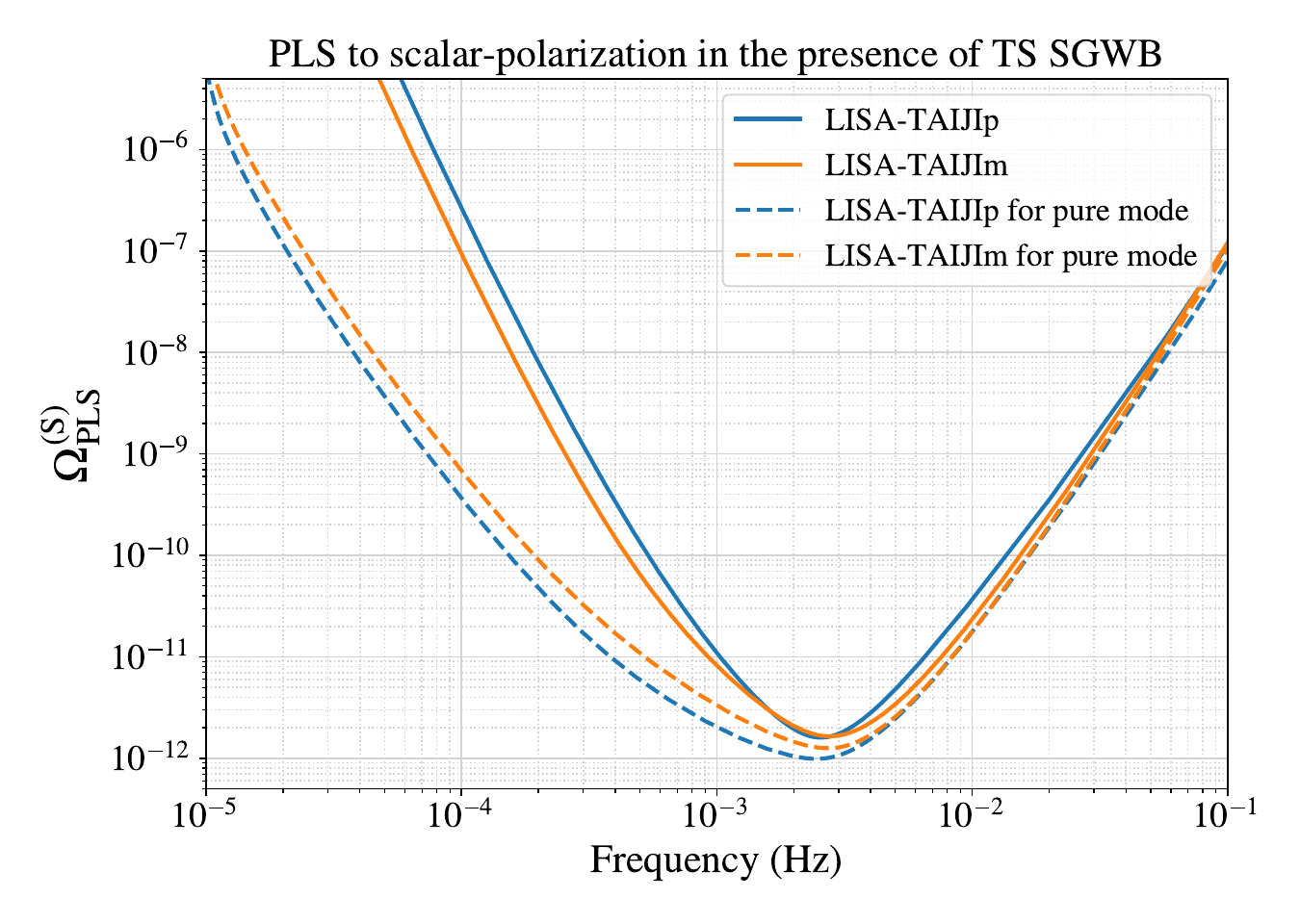}
\caption{\label{fig:sensitivity_LISA_TAIJI_TS} Sensitivities of the LISA-TAIJI detector networks to the tensor (top row) and scalar (bottom row) polarization components of a SGWB containing tensor and scalar modes. The left panels show the effective fractional energy-density sensitivities, $\Omega_{\rm eff}^{(P)}(f)$, while the right panels display the corresponding PLSs, $\Omega_{\rm PLS}^{(P)}$, assuming a detection threshold of $\rho_{\rm th}=10$ and an observation time of $T_{\rm obs}=3$ yr. Blue and orange curves correspond to the LISA-TAIJIp and LISA-TAIJIm configurations, respectively. Dashed curves denote the pure-polarization sensitivities, while solid curves show the sensitivities obtained after simultaneously fitting both tensor and scalar polarization components.}
\end{figure*}

\begin{figure*}[htbp]
\includegraphics[width=0.48\textwidth]{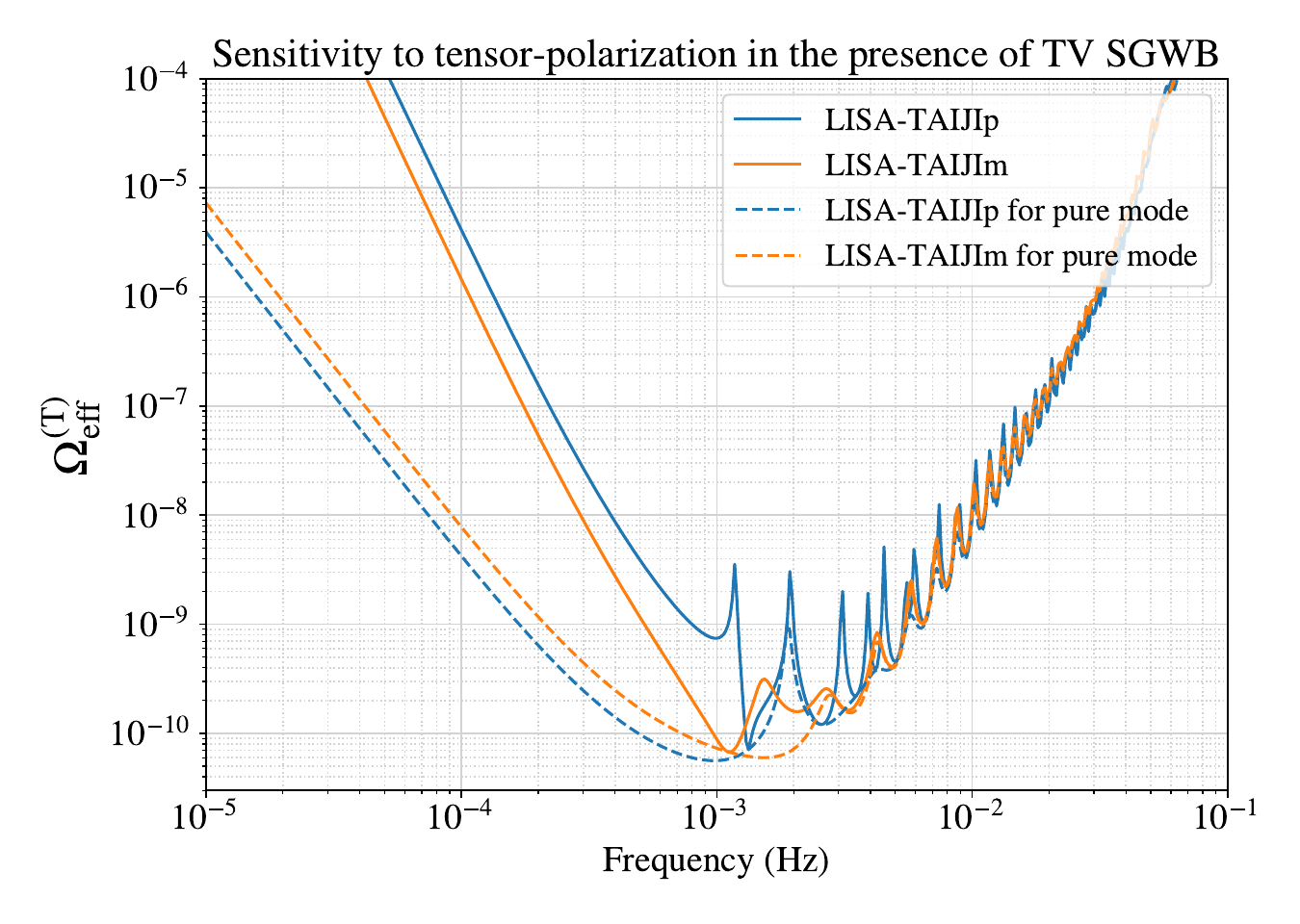} 
\includegraphics[width=0.48\textwidth]{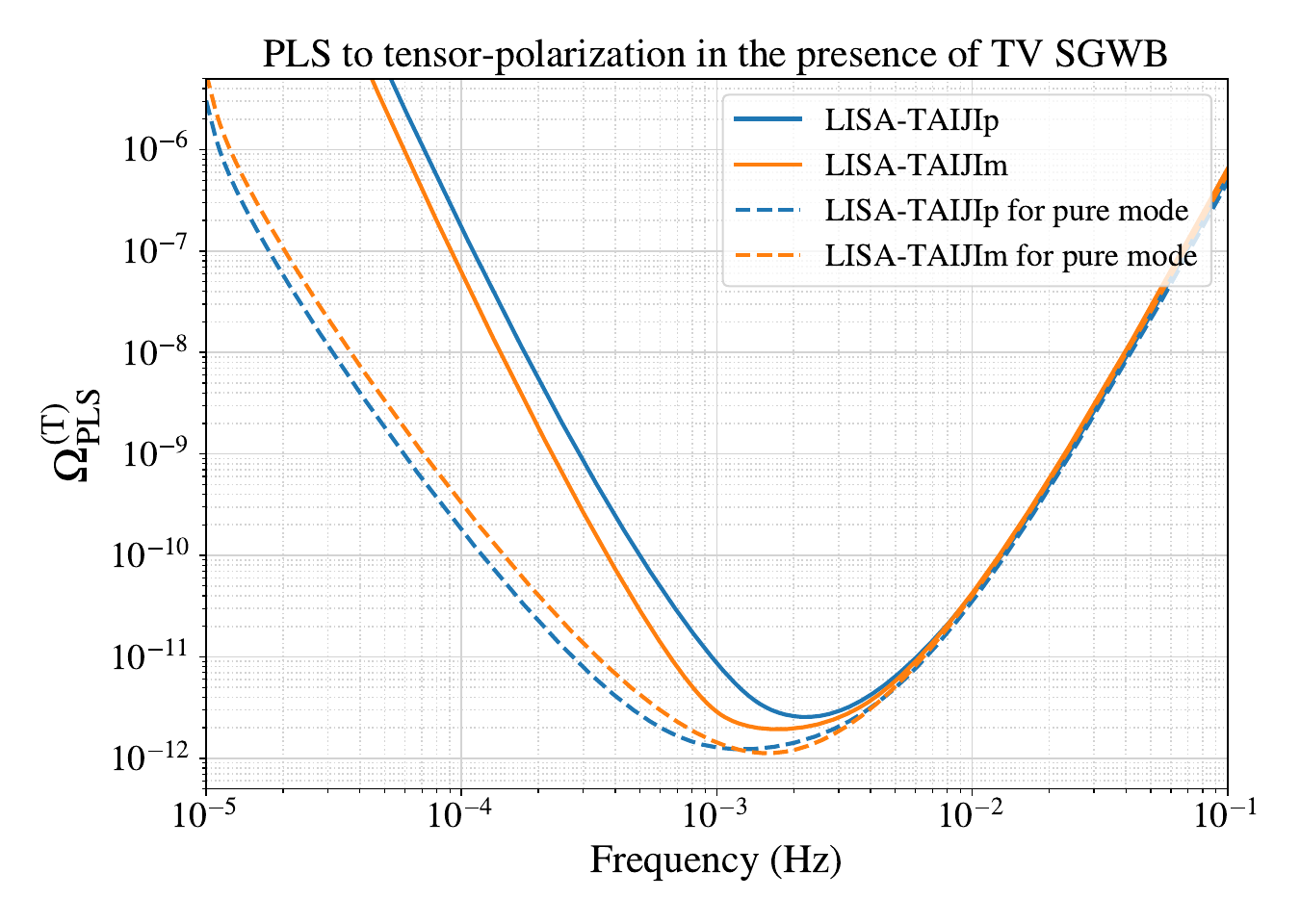}
\includegraphics[width=0.48\textwidth]{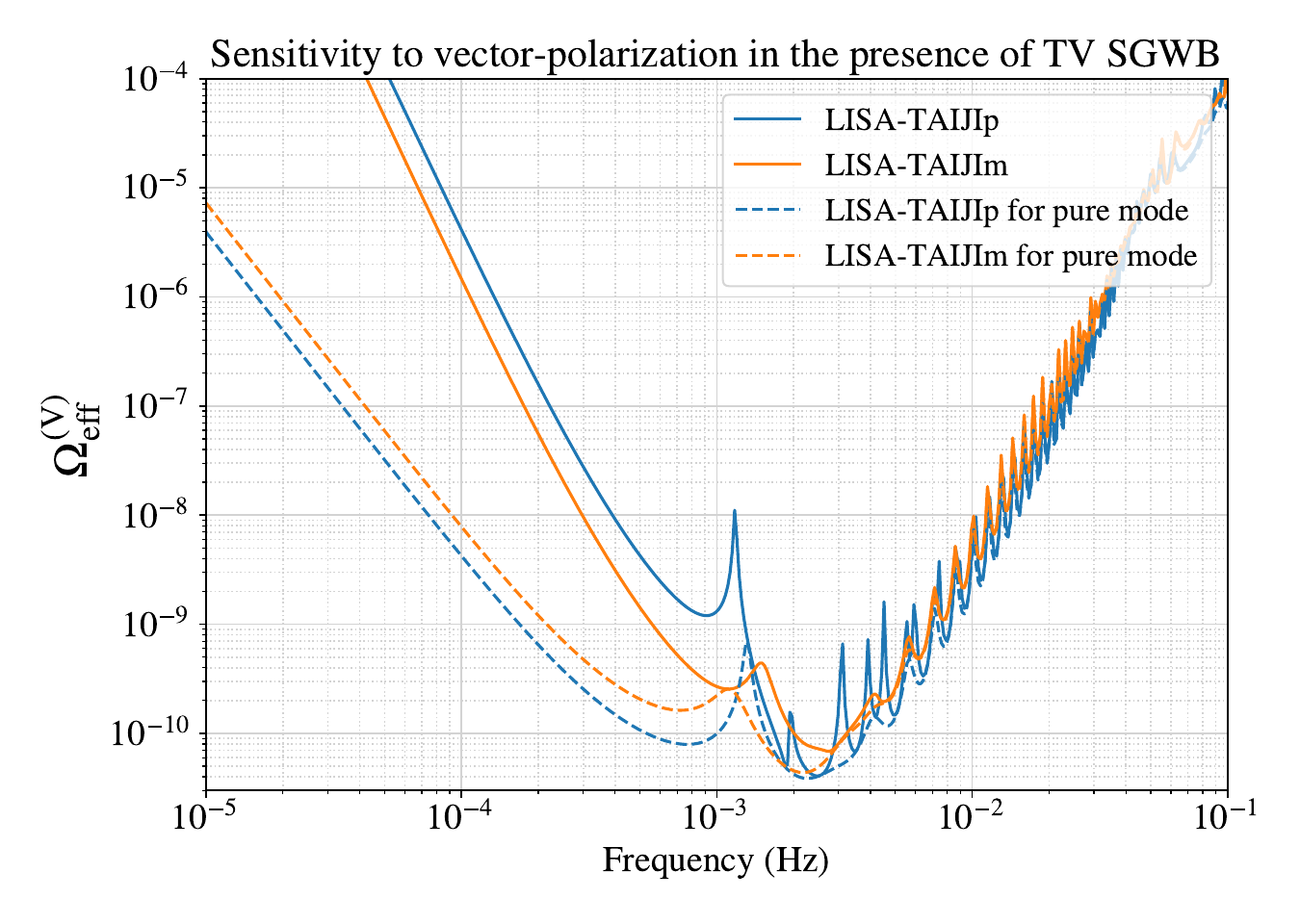}
\includegraphics[width=0.48\textwidth]{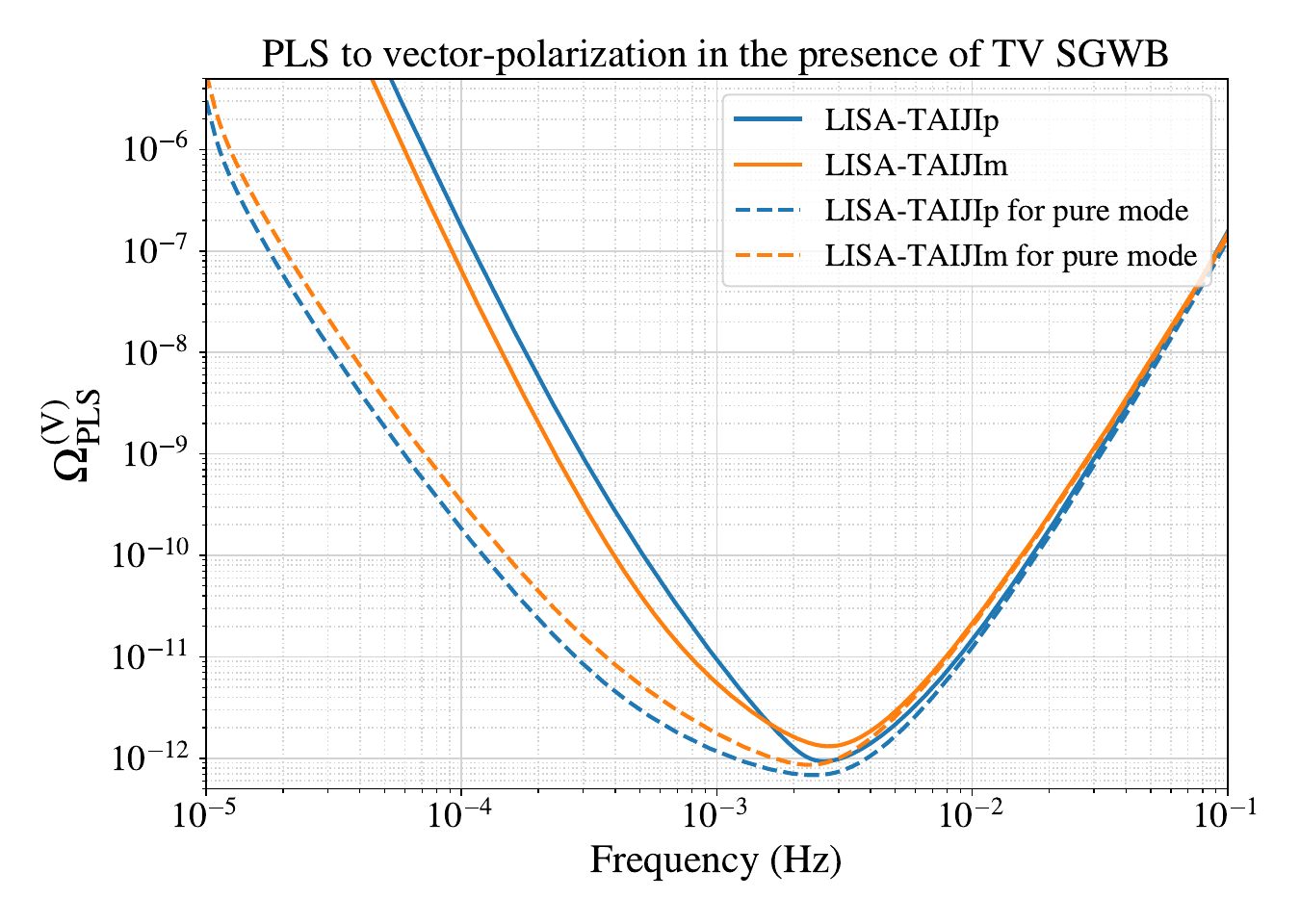}
\caption{ \label{fig:sensitivity_LISA_TAIJI_TV} Sensitivities of the LISA-TAIJI detector networks to the tensor (top row) and vector (bottom row) polarization components of a SGWB containing tensor and vector modes. The left panels show the effective fractional energy-density sensitivities, $\Omega_{\rm eff}^{(P)}(f)$, while the right panels display the corresponding PLSs, $\Omega_{\rm PLS}^{(P)}$, assuming a detection threshold of $\rho_{\rm th}=10$ and an observation time of $T_{\rm obs}=3$ yr. Blue and orange curves correspond to the LISA-TAIJIp and LISA-TAIJIm configurations, respectively. Dashed curves denote the pure-polarization sensitivities, while solid curves show the sensitivities obtained after simultaneously fitting both tensor and vector polarization components.}
\end{figure*}

\begin{figure*}[htbp]
\includegraphics[width=0.48\textwidth]{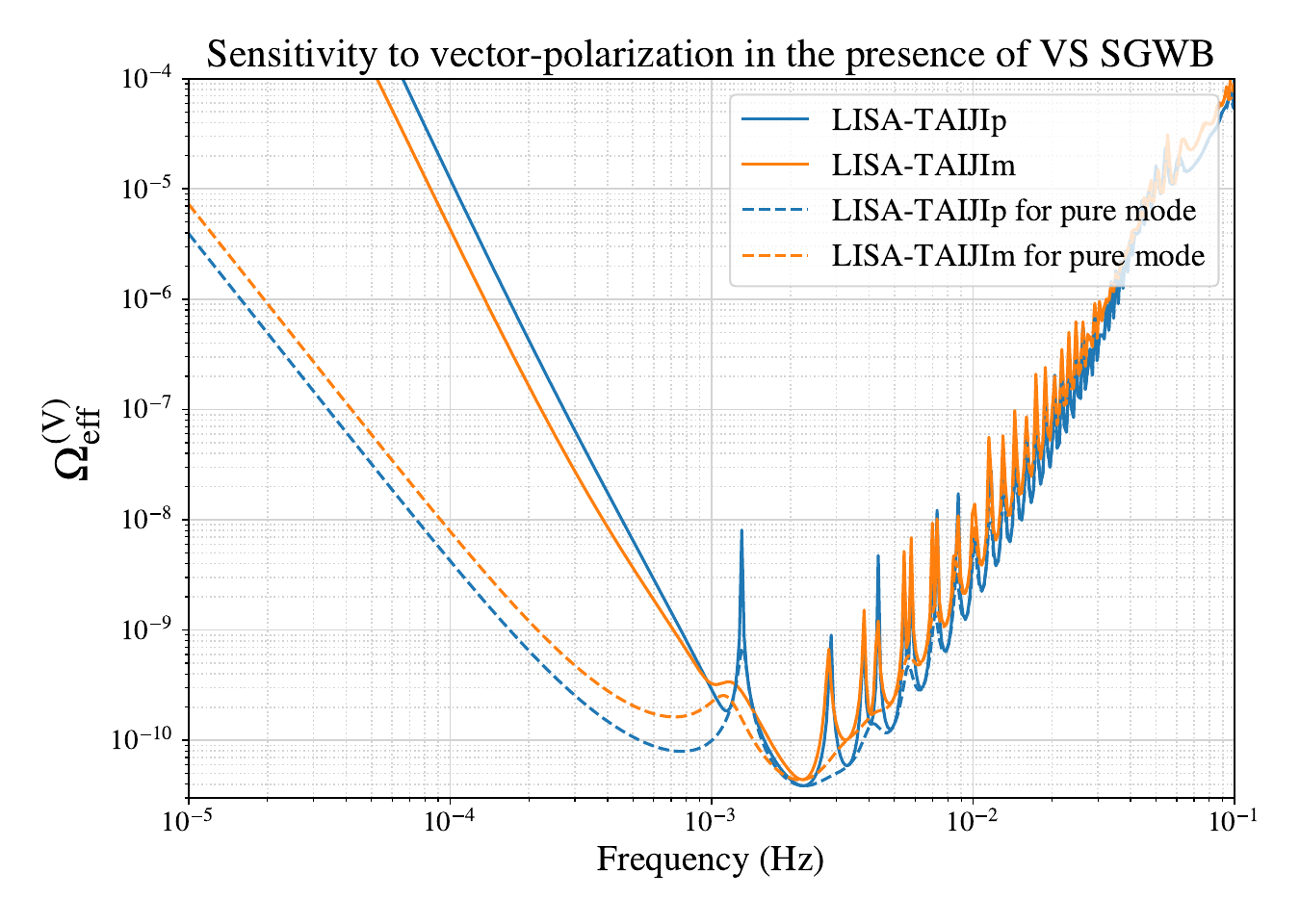}
\includegraphics[width=0.48\textwidth]{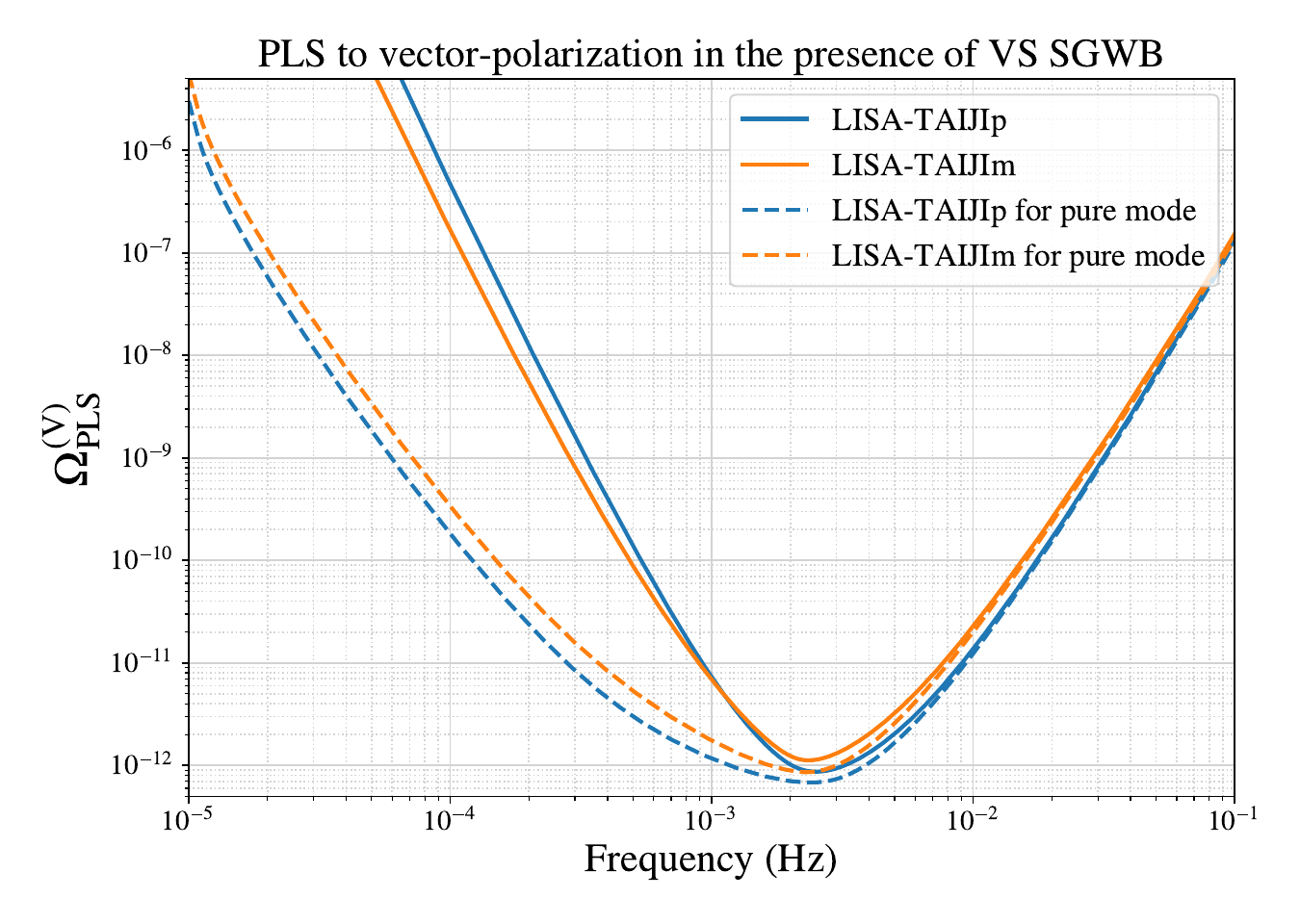}
\includegraphics[width=0.48\textwidth]{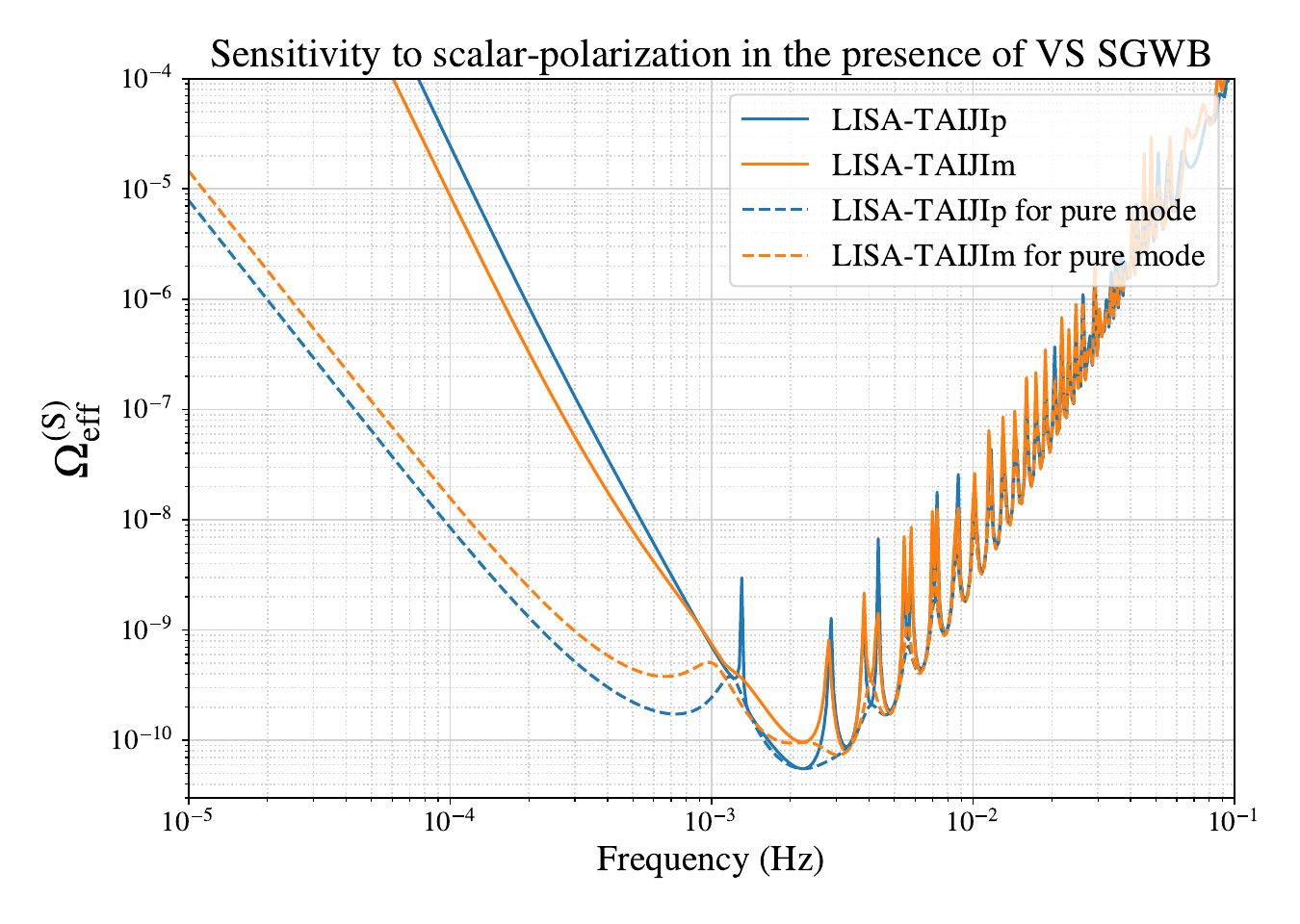} 
\includegraphics[width=0.48\textwidth]{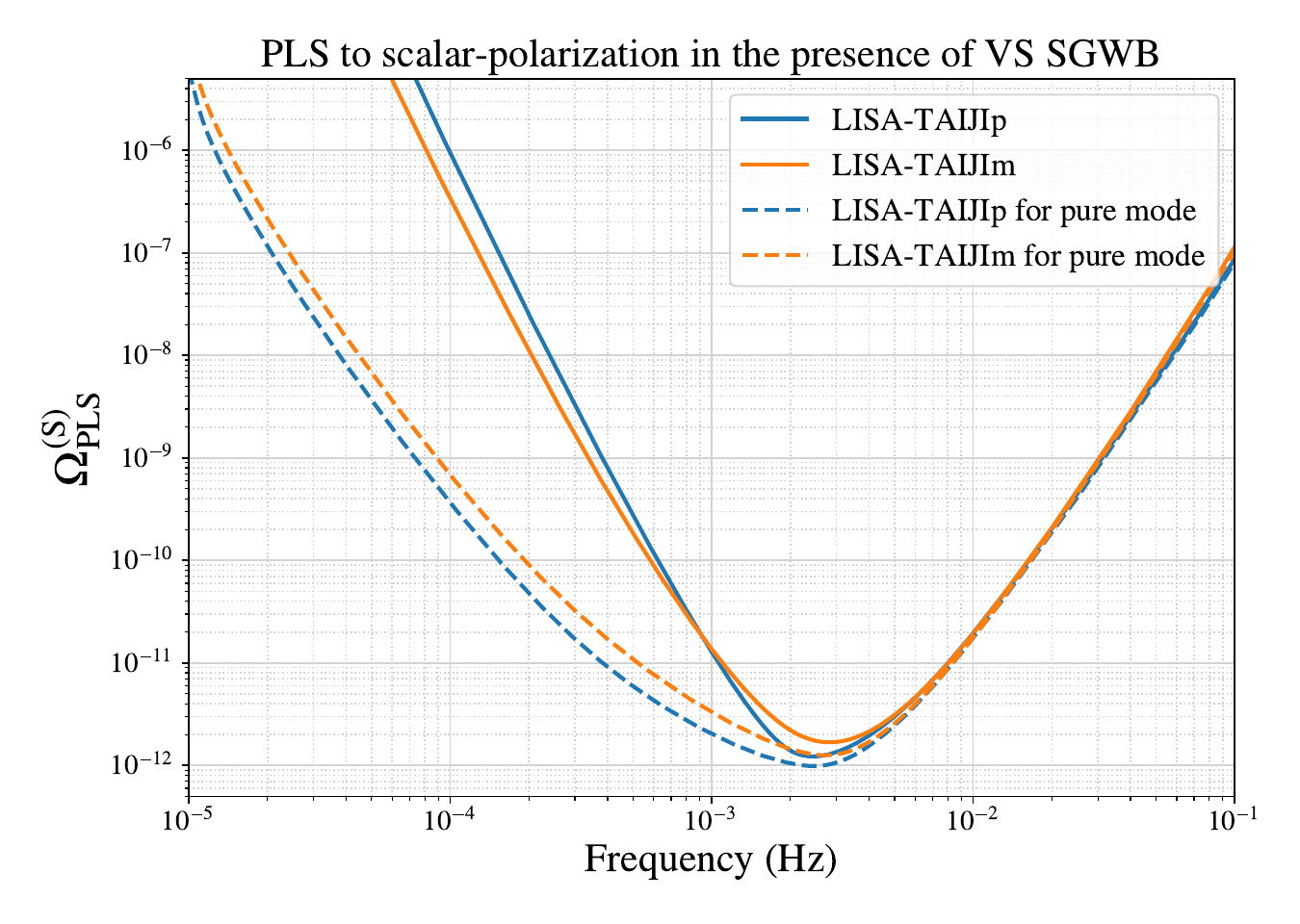}
\caption{ \label{fig:sensitivity_LISA_TAIJI_VS} Sensitivities of the LISA-TAIJI detector networks to the vector (top row) and scalar (bottom row) polarization components of a SGWB containing vector and scalar modes. The left panels show the effective fractional energy-density sensitivities, $\Omega_{\rm eff}^{(P)}(f)$, while the right panels display the corresponding PLSs, $\Omega_{\rm PLS}^{(P)}$, assuming a detection threshold of $\rho_{\rm th}=10$ and an observation time of $T_{\rm obs}=3$ yr. Blue and orange curves correspond to the LISA-TAIJIp and LISA-TAIJIm configurations, respectively. Dashed curves denote the pure-polarization sensitivities, while solid curves show the sensitivities obtained after simultaneously fitting both vector and scalar polarization components.}
\end{figure*}

To evaluate the detection performance of the LISA-TAIJI network, we analyze the sensitivities to different polarization modes (tensor, vector, and scalar) under four SGWB scenarios: a mixed tensor-vector-scalar (TVS) background (Fig. \ref{fig:sensitivity_LISA_TAIJI_TVS}), a TS background (Fig. \ref{fig:sensitivity_LISA_TAIJI_TS}), a TV background (Fig. \ref{fig:sensitivity_LISA_TAIJI_TV}), and a vector-scalar (VS) background (Fig. \ref{fig:sensitivity_LISA_TAIJI_VS}). The effective energy density sensitivities $\Omega^{(P)}_{\text{eff}}(f)$ are presented in the left panels, while the corresponding PLS, $\Omega^{(P)}_{\text{PLS}}(f)$, are illustrated in the right panels. 
In all scenarios, the solid curves, which account for the simultaneous presence of multiple polarization sectors, lie above the corresponding dashed curves obtained under the pure-mode assumption. This behavior reflects the additional statistical uncertainty introduced by polarization separation, where correlations among different polarization responses degrade the sensitivity to any individual component. The effect is particularly pronounced at low frequencies, where the responses of different polarization sectors become more difficult to distinguish.

The strongest degradation occurs in the TVS scenario shown in Fig.~\ref{fig:sensitivity_LISA_TAIJI_TVS}. When tensor, vector, and scalar components are simultaneously included in the signal model, the component-separation problem becomes three-dimensional and involves the largest number of covariance terms. As a result, the effective sensitivities deviate substantially from their pure-mode limits. This degradation is especially significant for the LISA-TAIJIp configuration, whose mixed-mode sensitivities differ considerably from the corresponding pure-mode sensitivities. By contrast, the degradation is noticeably weaker for LISA-TAIJIm, indicating that its larger relative inclination improves the linear independence of the network responses and enhances the separation of different polarization sectors.

When the number of simultaneously fitted polarization sectors is reduced from three to two, the sensitivity degradation becomes substantially weaker. This trend is evident in the TS, TV, and VS scenarios shown in Figs. ~\ref{fig:sensitivity_LISA_TAIJI_TS}-\ref{fig:sensitivity_LISA_TAIJI_VS}. Compared with the TVS case, the mixed-mode sensitivities remain closer to the pure-mode limits across most of the observational band. The reduction in dimensionality decreases the number of covariance terms in the component-separation matrix and alleviates the degeneracies among polarization sectors.
Interestingly, the relative performance of the two network configurations depends on the specific polarization combination. For the TS and TV scenarios, the LISA-TAIJIm configuration generally retains better sensitivities than LISA-TAIJIp, particularly below several millihertz. In contrast, the difference between the two networks becomes much smaller in the VS scenario and can even reverse over part of the frequency range, with LISA-TAIJIp exhibiting slightly better sensitivities for certain polarization components.

These results indicate that the advantage of the large inclination LISA-TAIJIm configuration is not simply associated with a stronger overall network sensitivity. Rather, it arises from the way in which the detector geometry modifies the correlations among different polarization responses. The improvement provided by LISA-TAIJIm is most pronounced in the TVS scenario and remains evident in the TS and TV cases, whereas it becomes weaker in the VS case. This behavior suggests that the impact of detector geometry on polarization separation depends sensitively on the specific combination of polarization sectors present in the SGWB.

Taken together, Figs.~\ref{fig:sensitivity_LISA_TAIJI_TVS}-\ref{fig:sensitivity_LISA_TAIJI_VS} show that the dominant loss of sensitivity arises when multiple polarization sectors coexist and must be simultaneously disentangled from the same set of cross-correlation measurements. While both LISA-TAIJI configurations achieve comparable sensitivities in the pure-mode limit, the LISA-TAIJIm network generally preserves a larger fraction of its intrinsic sensitivity when polarization separation is required. The degree of improvement, however, depends on the polarization content of the SGWB, demonstrating that detector geometry plays a central role in determining the effectiveness of polarization-component separation in future space-based detector networks.

\subsection{Detection performance for power-law SGWB}

The isotropic SGWB is conventionally characterized by its fractional energy density per logarithmic frequency interval,
\begin{equation}
\Omega_{\rm GW}(f)=\frac{1}{\rho_c}\frac{d\rho_{\rm GW}}{d\ln f},
\end{equation}
where $\rho_c=3H_0^2c^2/(8\pi G)$ denotes the critical energy density of the Universe \cite{Allen:1996vm,Allen:1997ad}. Such backgrounds may originate from both unresolved astrophysical populations and a variety of processes in the early Universe \cite{Caprini:2015zlo,Caprini:2019egz,Kuroyanagi:2018csn,Caprini:2019pxz,Flauger:2020qyi,Martinovic:2020hru,Pi:2022zxs,LiGong:2024qmt}. To assess the detection performance of the LISA-TAIJI networks, we adopt a fiducial power-law spectrum,
\begin{equation}
\Omega^{(P)}_\mathrm{GW} = A_P \left( \frac{f}{f_\mathrm{ref}} \right)^{\alpha_P}, \label{eq:PL_signal} 
\end{equation}
where $A_P$ and $\alpha_P$ denote the amplitude and spectral index of polarization component $P$, respectively.
Throughout this analysis, we use a tensor background with $A_T=4.446\times10^{-12}$ and $\alpha_T=2/3$ at the reference frequency $f_{\rm ref}=1~{\rm mHz}$ as the fiducial signal \cite{LIGOScientific:2019vic}. The corresponding strain PSD is related to the fractional energy density through
\begin{equation} 
S^{(P)}_h (f)  = \frac{3 H^2_0}{4 \pi^2 f^3} \Omega^{(P)}_\mathrm{GW} (f). \label{eq:P_h}
\end{equation}

To quantify the impact of polarization degeneracies on SGWB detection, we evaluate the SNR after marginalizing over the remaining polarization components. The resulting profiled SNR for polarization mode $P$ is given by
\begin{equation}
 \rho_{(P)}^2 =  \sum_{i,j \in \{\mathrm{A, E, T}\} } 2 T_\mathrm{obs} \int^{+ \infty }_{0}  \left[ \frac{ S_h^{(P)} (f)  }{ \sigma_P (M (f)) } \right]^2 \mathrm{d} f , 
\end{equation}
where $M(f)$ is defined in Eq.~\eqref{eq:M_factor} and includes contributions from both instrumental noise and SGWB signals. In the weak-signal limit, $M(f)$ reduces to the product of the noise spectra, whereas the full expression is retained when the SGWB contribution is non-negligible.
The SNR calculation employs the amplitude $A_P$ in Eq. \eqref{eq:PL_signal} as free parameters, while keeping the power index fixed at $\alpha_P =2/3$ across all polarization modes. The profiled SNR is obtained by maximizing the likelihood over the remaining polarization amplitudes treated as nuisance parameters, thereby projecting the higher-dimensional parameter space onto two-dimensional subspaces. To facilitate a direct comparison between the two network configurations, we further evaluate the local SNR ratio, $\text{SNR}_{\rm TAIJIp} / \text{SNR}_{\rm TAIJIm}$, across the parameter space spanned by the polarization amplitudes.

\begin{figure*}[htbp]
    \centering
    \includegraphics[width=0.88\textwidth]{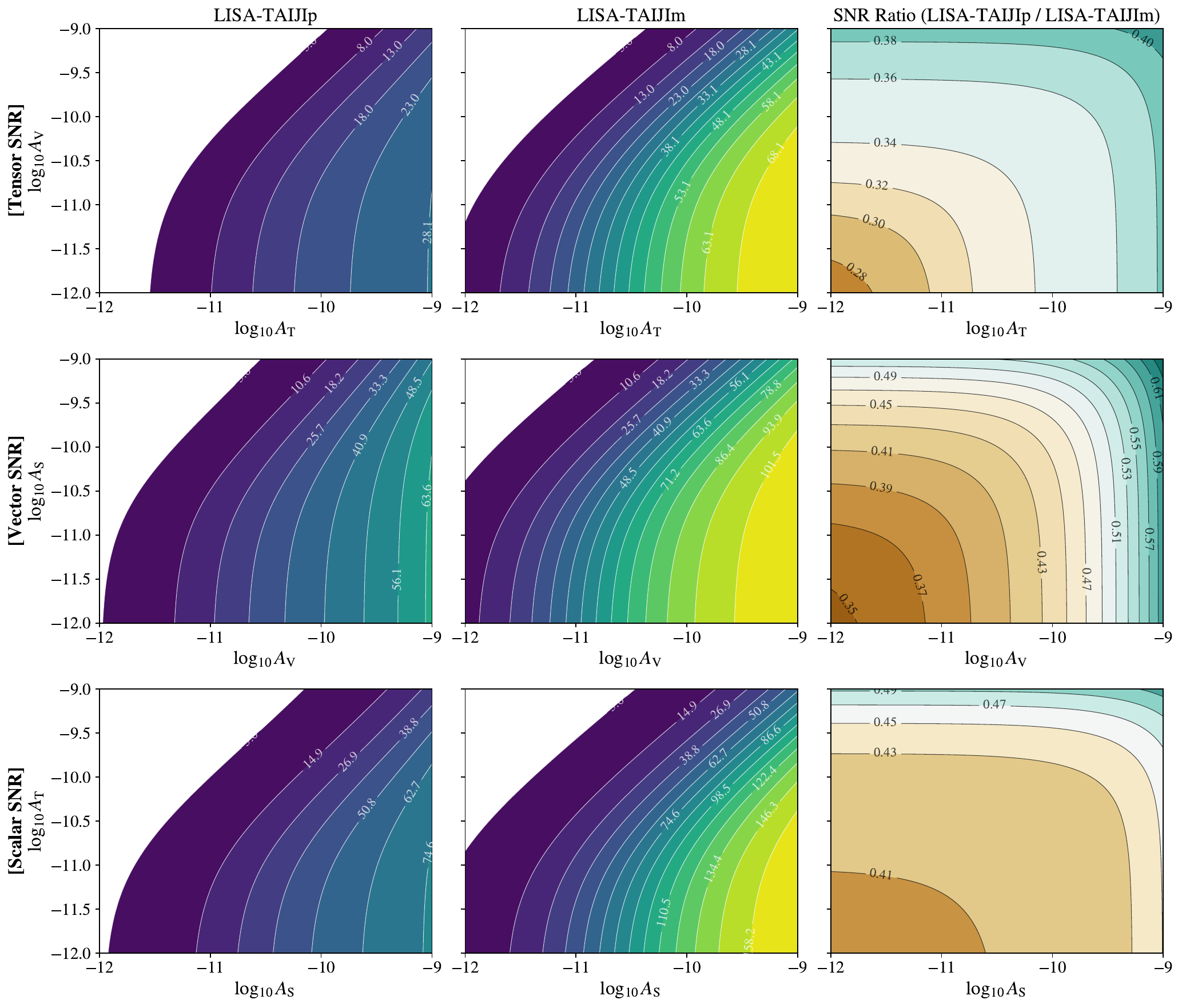}
    \caption{The profiled SNRs for the TVS SGWB. The three rows correspond to the recovered tensor, vector, and scalar polarization components, respectively. For each row, the horizontal axis represents the amplitude of the target polarization component, while the vertical axis denotes the amplitude of one of the remaining polarization sectors. The left and middle columns show the profiled SNRs obtained with the LISA-TAIJIp and LISA-TAIJIm configurations, respectively, and the right column presents the corresponding SNR ratio, $\text{SNR}_{\rm LISA-TAIJIp} / \text{SNR}_{\rm LISA-TAIJIm}$. The ratio contours indicate that the LISA-TAIJIm configuration consistently achieves higher effective SNRs, with improvements reaching factors of $\sim$2-3.
}
    \label{fig:snr_ratio_TVS}
\end{figure*}

Figure~\ref{fig:snr_ratio_TVS} shows the profiled SNRs for the TVS scenario, where all three polarization sectors are simultaneously present and should be separated from the same set of cross-correlation measurements. The three rows correspond to the recovered tensor, vector, and scalar polarization components, respectively. For each row, the horizontal axis denotes the amplitude of the target polarization component, while the vertical axis corresponds to the amplitude of one of the remaining polarization sectors, with the third polarization mode profiled over in the likelihood analysis.

A common trend is observed across all three polarization sectors. The recovered SNR increases with the amplitude of the target polarization component, reflecting the larger contribution of the corresponding SGWB signal to the cross-correlation measurements. Conversely, the recovered SNR decreases as the amplitude of the competing polarization component increases. Although the additional polarization component also contributes signal power, it simultaneously enhances the covariance among different polarization sectors and therefore reduces the ability of the detector network to isolate the target mode. This behavior reflects the impact of polarization degeneracies in the component-separation analysis.

Comparing the first and second columns in Fig. \ref{fig:snr_ratio_TVS}, the LISA-TAIJIm configuration consistently achieves larger profiled SNRs throughout the parameter space. This difference is quantified in the right column, which displays the ratio ${\rm SNR}_{\rm LISA\text{-}TAIJIp}/{\rm SNR}_{\rm LISA\text{-}TAIJIm}$. For all three polarization sectors, the ratio remains within approximately $0.3$-$0.6$, indicating that the recovered SNR of the LISA-TAIJIp network is significantly reduced relative to that of LISA-TAIJIm. These results are consistent with the sensitivity analysis presented in Fig.~\ref{fig:sensitivity_LISA_TAIJI_TVS}, where the LISA-TAIJIm configuration exhibited a smaller degradation from the pure-mode sensitivities. The improved performance of LISA-TAIJIm can be attributed to its larger relative inclination, which enhances the complementarity of the ORFs and mitigates parameter degeneracies among the tensor, vector, and scalar polarization sectors.

\begin{figure*}[htbp]
    \centering
    \includegraphics[width=0.88\textwidth]{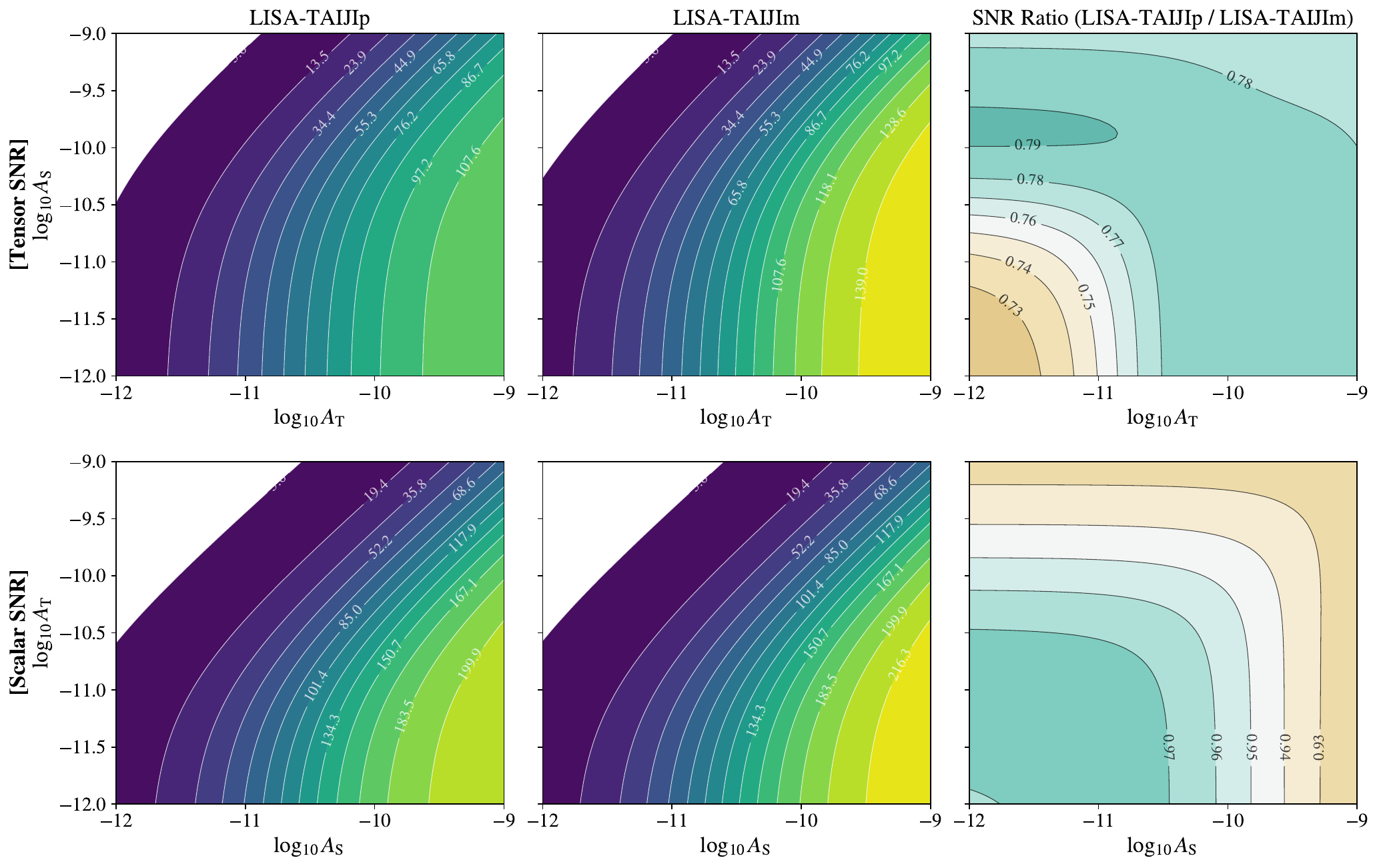}
    \caption{Profiled SNRs for a TS SGWB. The top and bottom rows correspond to the recovered tensor and scalar polarization components, respectively. The LISA-TAIJIp configuration exhibits a smaller reduction in SNR, with the ratio remaining within approximately $0.73$-$0.79$ for the tensor mode and $0.93$-$0.97$ for the scalar mode.
    }
    \label{fig:snr_ratio_TS}
\end{figure*}

\begin{figure*}[htbp]
    \centering
    \includegraphics[width=0.88\textwidth]{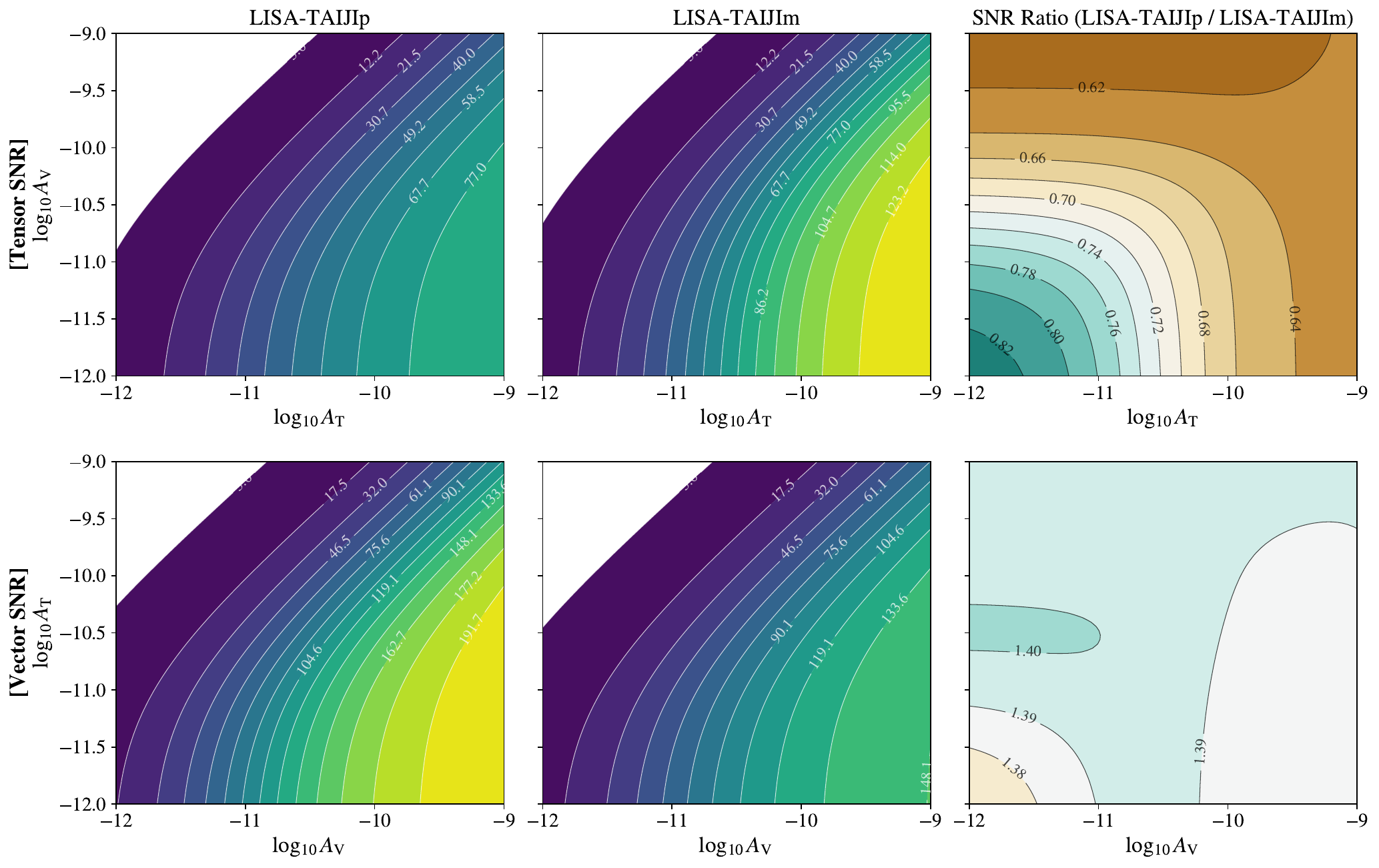}
    \caption{Profiled SNRs for a TV SGWB. The top and bottom rows correspond to the recovered tensor and vector polarization components, respectively. For the vector mode, the SNR ratio reaches approximately 1.4, indicating that the LISA-TAIJIp configuration can achieve a higher SNR for this particular polarization sector over current parameter space.}
    \label{fig:snr_ratio_TV}
\end{figure*}

\begin{figure*}[htbp]
    \centering
    \includegraphics[width=0.88\textwidth]{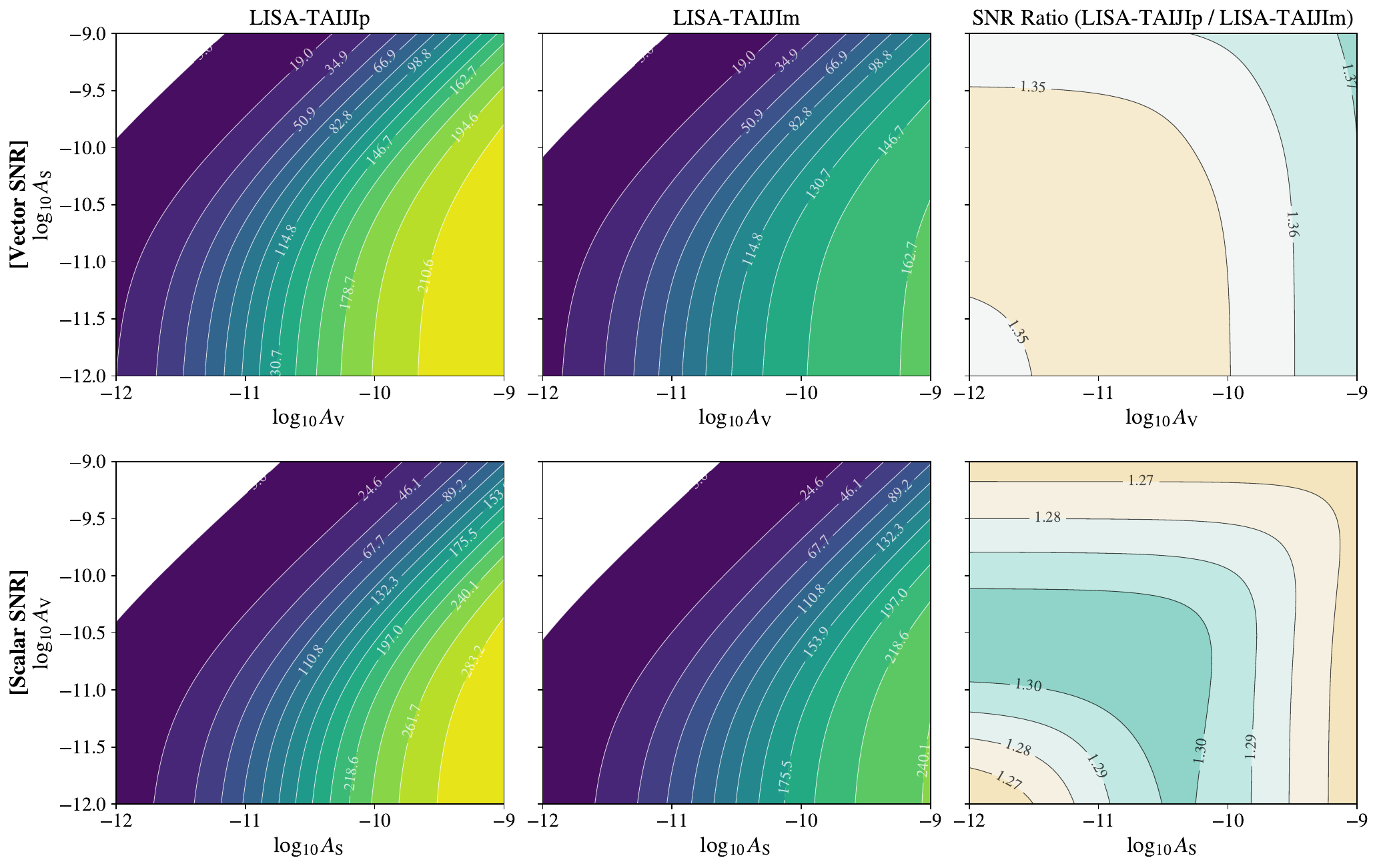}
    \caption{Profiled SNRs for a VS SGWB. The top and bottom rows correspond to the recovered vector and scalar polarization components, respectively. Unlike the TVS, TS, and TV cases, the LISA-TAIJIp configuration consistently achieves higher recovered SNRs for both polarization components. The ratio reaches approximately $1.35$--$1.37$ for the vector mode and $1.27$--$1.30$ for the scalar mode.
    }
    \label{fig:snr_ratio_VS}
\end{figure*}

As the dimensionality of the confusion background is reduced from three polarization sectors to two, the impact of parameter covariance becomes substantially weaker, leading to a noticeable change in the relative performance of the two network configurations. Figures~\ref{fig:snr_ratio_TS}, \ref{fig:snr_ratio_TV}, and \ref{fig:snr_ratio_VS} present the profiled SNRs and their corresponding ratios for the TS, TV, and VS backgrounds, respectively.
Compared with the TVS case, the LISA-TAIJIp configuration experiences a much smaller loss of SNR in all dual-polarization scenarios. For the tensor component, the SNR ratio remains predominantly within 0.73-0.79 in the TS scenario and 0.62-0.82 in the TV scenario. These values indicate that the advantage of LISA-TAIJIm is significantly reduced once the dimensionality of the polarization-separation problem decreases.
The relative performance depends strongly on the polarization combination. In the TS scenario, the scalar component achieves nearly identical performance in the two networks, with the ratio reaching 0.93-0.97. In the TV scenario, the vector component becomes more detectable in the LISA-TAIJIp network, yielding a ratio of approximately 1.38-1.40. This behavior is consistent with the sensitivity results shown in Fig.~\ref{fig:sensitivity_LISA_TAIJI_TV}, where LISA-TAIJIp exhibits superior sensitivity to the vector polarization over the frequency range of roughly 2-20 mHz.

A similar trend is observed in the VS scenario shown in Fig.~\ref{fig:snr_ratio_VS}. In this case, the LISA-TAIJIp configuration consistently achieves higher profiled SNRs for both polarization sectors across the entire parameter space. The SNR ratio reaches approximately 1.3 for both the vector and scalar components, indicating a systematic advantage of LISA-TAIJIp over LISA-TAIJIm. This result contrasts with the behavior observed in the TVS scenario and demonstrates that the superiority of the LISA-TAIJIm configuration is not universal.
A likely explanation is that the adopted power-law spectrum with spectral index $\alpha=2/3$ places relatively greater weight on the higher-frequency portion of the observational band. As shown in Fig.~\ref{fig:sensitivity_LISA_TAIJI_VS}, the LISA-TAIJIp configuration exhibits better sensitivities to both vector and scalar polarizations over a broad range of intermediate and high frequencies. Consequently, these frequencies contribute more significantly to the SNR integral, leading to systematically larger recovered SNRs for LISA-TAIJIp in the VS scenario.

Taken together, the TS, TV, and VS results suggest that the benefit of the large-inclination LISA-TAIJIm configuration is most pronounced when tensor modes are simultaneously present with additional polarization sectors. In contrast, when only vector and scalar modes are considered, the LISA-TAIJIp network can achieve comparable or even superior performance. Combined with the sensitivity results presented in Sec. \ref{sec:separation_sensitivity}, these findings indicate that the primary role of detector geometry is to reduce the degeneracies associated with tensor-containing polarization mixtures, whereas its impact on VS separation is considerably weaker.

\section{Forecasting constraints on SGWB components} \label{sec:FIM_forecast}

In the previous section, we investigated the capability of the LISA-TAIJI networks to separate different SGWB polarization components in a largely model-independent manner. We now consider a complementary problem: constraining the parameters of stochastic backgrounds with prescribed spectral models. Specifically, we employ a Fisher information matrix (FIM) analysis to evaluate the expected constraints on the amplitudes and spectral indices of power-law SGWB components. In contrast to previous studies \cite[and references therein]{Boileau:2020rpg}, only cross-correlations between the two missions are included in the analysis, while auto-correlations among the TDI channels within each mission are excluded. This choice isolates the contribution of the network geometry and enables a direct comparison between the LISA-TAIJIp and LISA-TAIJIm configurations.

\subsection{Fisher matrix analysis formalism}

To forecast parameter constraints on mixed SGWB models, we assume that each polarization component is described by a power-law spectrum and evaluate the corresponding Fisher information matrix. Unlike the polarization-separation analysis presented in previous section, where tensor, vector, and scalar components are treated as unknown contributions to be reconstructed directly from the data, the present analysis assumes that the polarization sectors are already incorporated into the signal model. The Fisher matrix therefore quantifies the precision with which the model parameters can be measured, rather than the ability of the detector network to identify different polarization sectors directly from the observations.

The SGWB contribution to the cross-correlation spectrum between two TDI channels can be written as
\begin{equation}
\begin{aligned}
\mathcal{S}_{h, \mathrm{tot}} = &  \sum_{P} \Gamma^{(P)}_{ij} (f) S^{(P)}_h (f) \\
 = & \left(\frac{3 H^2_0}{ 4 \pi ^2 f^3} \right) \sum_{P} \Gamma^{(P)}_{ij} (f) \Omega^{(P)}_\mathrm{GW} (f) \\
 = & \left(\frac{3 H^2_0}{ 4 \pi ^2 f^3} \right) \sum_{P} \Gamma^{(P)}_{ij} (f) A_P \left( \frac{f}{f_\mathrm{ref}} \right)^{\alpha_P}, 
\end{aligned}
\end{equation}
The parameters to be estimated are the logarithmic amplitude $\log_{10}A_P$ and the spectral index $\alpha_P$ of each polarization component. And the partial derivatives of SGWB with respect to these parameters are
\begin{align}
\frac{\partial \mathcal{S}_{h, \mathrm{tot}} }{ \partial \log_{10} A_P  }  = & \left(\frac{3 H^2_0}{ 4 \pi ^2 f^3} \right) \Gamma^{(P)}_{ij} \frac{\partial \Omega^{(P)}_\mathrm{GW} }{ \partial \log_{10} A_P  } \notag \\
  = &  \left(\frac{3 H^2_0}{ 4 \pi ^2 f^3} \right) \Gamma^{(P)}_{ij}  \Omega^{(P)}_\mathrm{GW} ~ \ln 10, \label{eq:partial_lnA} \\
\frac{\partial \mathcal{S}_{h, \mathrm{tot}} }{ \partial \alpha_P }  = & \left(\frac{3 H^2_0}{ 4 \pi ^2 f^3} \right) \Gamma^{(P)}_{ij} \frac{\partial \Omega^{(P)}_\mathrm{GW} }{ \partial \alpha_P } \notag \\
  = &  \left(\frac{3 H^2_0}{ 4 \pi ^2 f^3} \right) \Gamma^{(P)}_{ij}  \Omega^{(P)}_\mathrm{GW} ~ \ln \frac{f}{ 1 \ \mathrm{mHz} }. \label{eq:partial_alpha}
\end{align}
It is worth noting that Eqs.~\eqref{eq:partial_lnA} and \eqref{eq:partial_alpha} involve only the response and spectrum of the target polarization component. Therefore, the Fisher derivatives do not require the reconstruction of mixed polarization signals and are not directly affected by polarization-separation degeneracies.
The FIM for parameters $\theta_a$ and $\theta_b$ will be
\begin{equation} \label{eq:FIM_cross}
\begin{aligned}
 \mathcal{F}_{ab } = & \sum_{i,j = \mathrm{A, E, T} } 2 T_\mathrm{obs} \int^{f_\mathrm{max}}_{0}  \frac{  \frac{\partial \mathcal{S}_{h, \mathrm{tot}}  }{ \partial \theta_a  }  \frac{\partial \mathcal{S}_{h, \mathrm{tot}} }{ \partial \theta_b  } }{ M(f)  } \mathrm{d} f \\ 
\end{aligned}
\end{equation}
The variance-covariance matrix of the parameters will be
\begin{equation}
\begin{aligned}
 \left\langle \delta \theta_i \delta \theta_j  \right\rangle = \left( \mathcal{F}^{-1}_{ab} \right)_{ij} + \mathcal{O}({\rho}^{-1})
\overset{{\rho} \gg 1}{\simeq } \left( \mathcal{F}^{-1}_{ab} \right)_{ij} . \\
\end{aligned}
\end{equation}
The standard deviation $\sigma_i$ of the parameter $\theta_i$ is
\begin{equation}
\begin{aligned}
\delta \theta_i & \overset{{\rho} \gg 1}{\simeq} \sqrt{ \left( \mathcal{F}^{-1}_{ab } \right)_{ii}  }. \\
\end{aligned}
\end{equation}

\subsection{Constraints on polarization parameters}

To assess the parameter-estimation capabilities of the two LISA-TAIJI networks, we consider four SGWB scenarios containing TVS, TS, TV, and VS polarization components. Each polarization sector is modeled by a power-law spectrum characterized by two parameters: the amplitude $\log_{10} A_P$ and the spectral index $\alpha_P$. The projected parameter constraints are obtained using the Fisher matrix formalism under the Gaussian approximation.

\begin{figure*}[htbp]
    \centering
    \includegraphics[width=\textwidth]{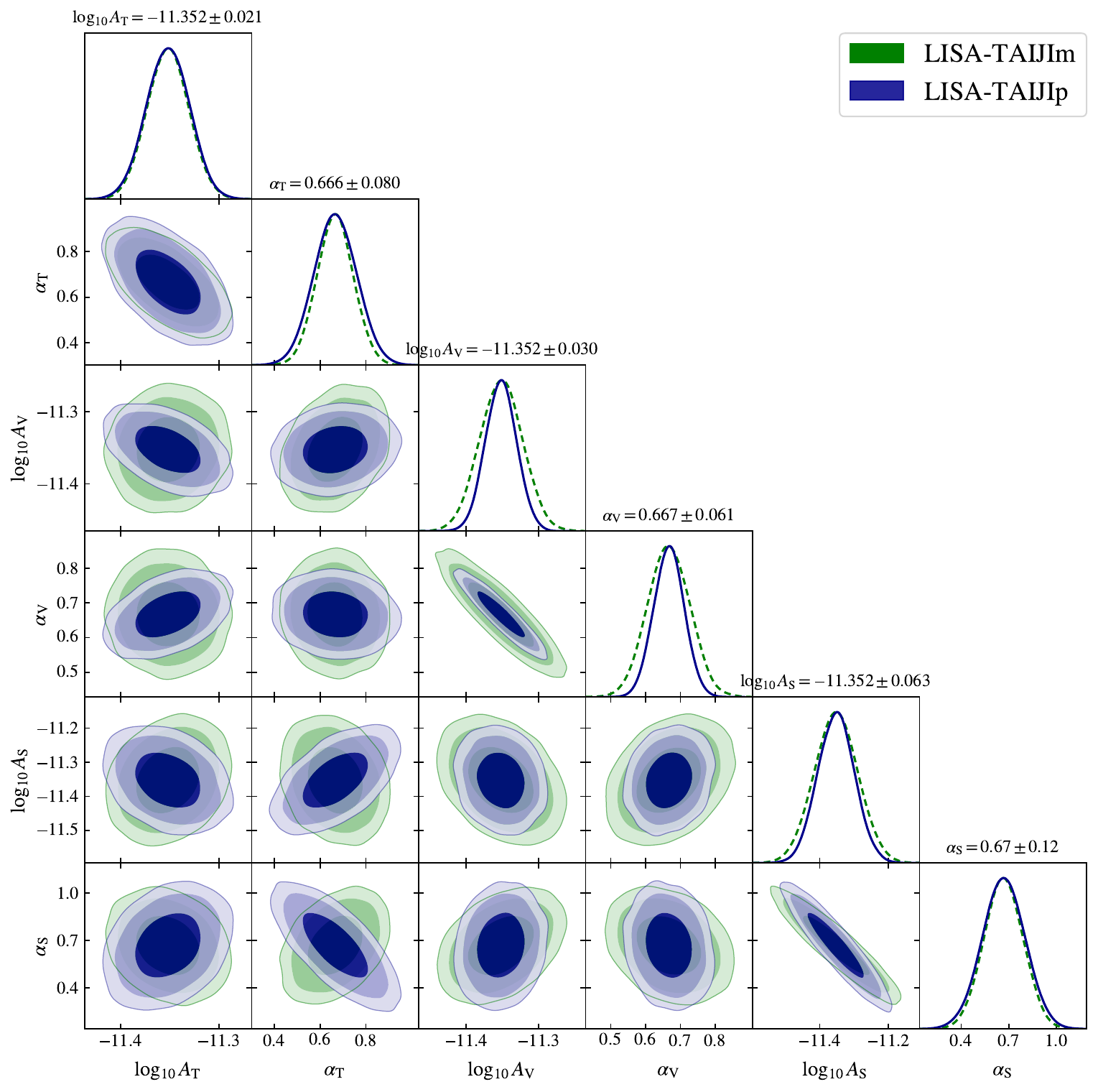}
    \caption{Fisher-forecasted parameter constraints for the TVS SGWB model. The six-dimensional parameter space consists of the amplitudes and spectral indices of the tensor, vector, and scalar polarization components. The contours represent the projected $1\sigma$, $2\sigma$, and $3\sigma$ confidence regions. Despite the substantial differences between the two network configurations in the polarization-separation analysis, the marginalized parameter constraints obtained with the LISA-TAIJIp (blue) and LISA-TAIJIm (green) configurations remain broadly comparable.}
    \label{fig:corner_TVS}
\end{figure*}

\begin{figure*}[htbp]
    \centering
    \includegraphics[width=0.7\textwidth]{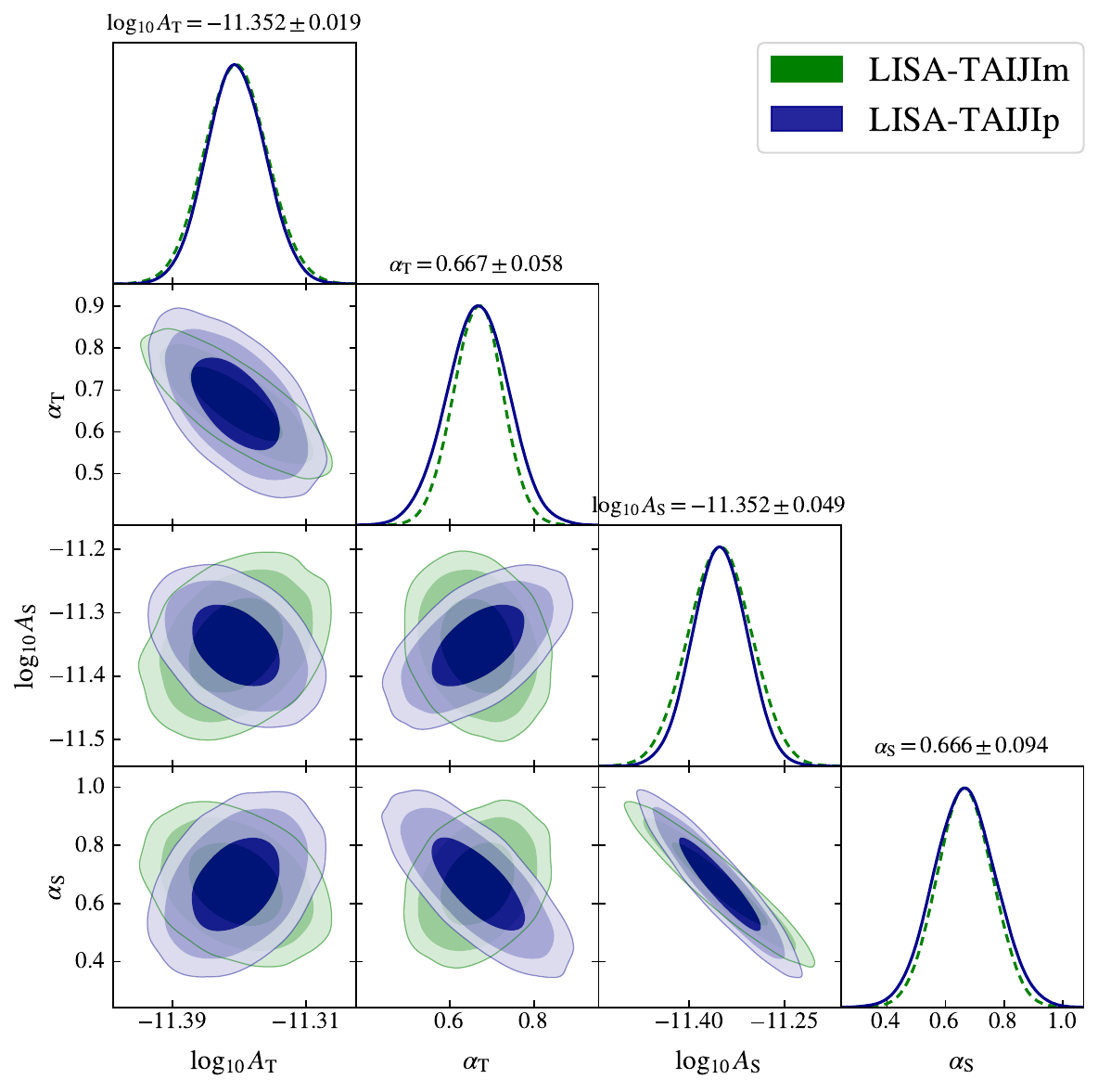}
    \caption{Fisher-forecasted parameter constraints for the TS SGWB model, encompassing four parameters (two amplitudes and two spectral indices). The LISA-TAIJIp network (blue) exhibits stronger parameter correlations (more tilted ellipses) than the LISA-TAIJIm network (green), reflecting its weaker ability to disentangle polarization modes due to a smaller inclination angle.}
    \label{fig:corner_TS}
\end{figure*}

\begin{figure*}[htbp]
    \centering
    \includegraphics[width=0.7\textwidth]{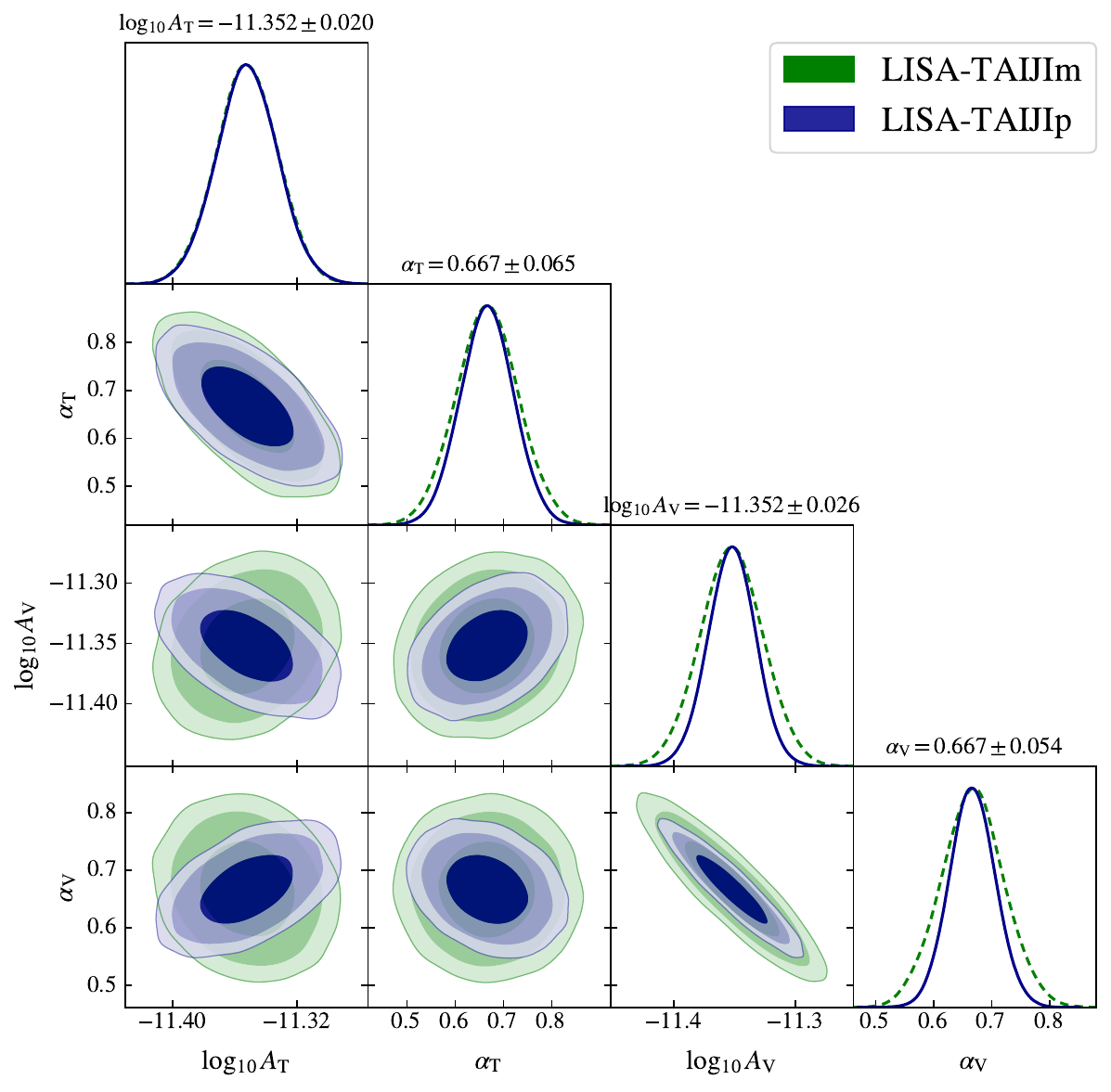}
    \caption{Fisher-forecasted parameter constraints for the TV SGWB scenario. Consistent with the other configurations, the strong parameter correlation in LISA-TAIJIp (blue) remains visible.}
    \label{fig:corner_TV}
\end{figure*}

Figures~\ref{fig:corner_TVS}, \ref{fig:corner_TS}, and \ref{fig:corner_TV} show the forecasted parameter constraints for the TVS, TS, and TV scenarios, respectively. A notable feature of all three figures is that the marginalized parameter uncertainties obtained with the LISA-TAIJIp and LISA-TAIJIm configurations are remarkably similar. This behavior contrasts with the sensitivity and SNR analyses presented in Sec.~\ref{sec:sensitivity_snr}, where the LISA-TAIJIm network exhibited substantially stronger capabilities for polarization separation, particularly in the TVS scenario.

Before discussing the origin of this difference, it is important to emphasize that the Fisher analysis addresses a fundamentally different statistical problem from the component-separation analysis. In Sec.~\ref{sec:sensitivity_snr}, the detector network is required to distinguish tensor, vector, and scalar contributions directly from the observed cross-correlation signals. The resulting sensitivities and SNRs therefore depend critically on the linear independence of the ORFs associated with different polarization sectors.

In contrast, the Fisher matrix is constructed from derivatives of a prescribed signal model. As shown in Eqs.~\eqref{eq:partial_lnA} and \eqref{eq:partial_alpha}, the derivatives with respect to $\log_{10}A_P$ and $\alpha_P$ involve only the response and spectrum of the corresponding polarization component. The tensor, vector, and scalar sectors are therefore already represented as distinct components within the signal model. Consequently, the Fisher analysis no longer addresses the problem of identifying different polarization sectors from the data, but instead quantifies how accurately the parameters of each component can be measured once these components are included in the model. As a result, the forecasted constraints are more closely related to the sensitivities of the individual polarization sectors than to the model-independent polarization-separation capability discussed in Sec.~\ref{sec:separation_sensitivity}. This substantially reduces the impact of the TVS degeneracies that dominate the sensitivity and SNR analyses.

Although the one-dimensional marginalized uncertainties are nearly identical for the two network configurations, systematic differences can still be identified in the two-dimensional confidence regions. In particular, the confidence ellipses associated with the LISA-TAIJIp configuration generally exhibit a larger tilt than those of LISA-TAIJIm, especially in Figs. \ref{fig:corner_TVS}-\ref{fig:corner_TV}. This behavior indicates stronger parameter correlations and is qualitatively consistent with the larger polarization degeneracies identified in the previous section. The origin of these correlations can be traced to the weaker geometrical separation of the ORFs in the LISA-TAIJIp configuration, which possesses a smaller relative inclination between the two detector constellations.

\begin{figure*}[htbp]
    \centering
    \includegraphics[width=0.7\textwidth]{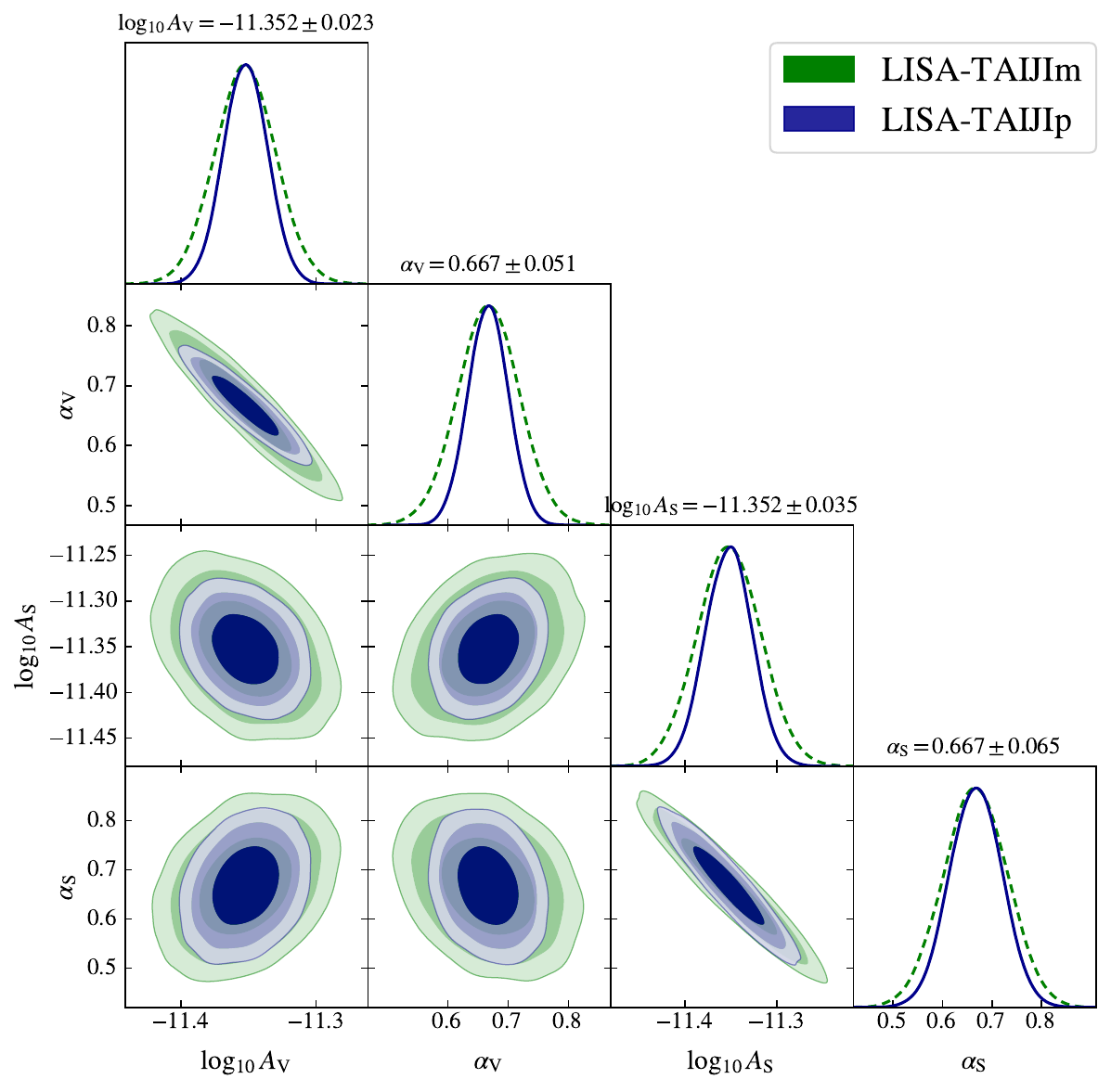}
    \caption{The parameter constraints for the VS SGWB scenario. The constraints obtained with the LISA-TAIJIp and LISA-TAIJIm configurations are nearly indistinguishable. Both the one-dimensional marginalized uncertainties and the two-dimensional confidence regions show only minor differences between the two network geometries.}
    \label{fig:corner_VS}
\end{figure*}

The VS scenario shown in Fig.~\ref{fig:corner_VS} exhibits a different behavior from the tensor-containing cases. Although the correlation structures of the two network configurations are broadly similar, as indicated by the comparable orientations and shapes of the confidence contours, the LISA-TAIJIp configuration yields systematically tighter constraints on both the amplitudes and spectral indices of the vector and scalar components than the LISA-TAIJIm configuration. The difference is therefore primarily reflected in the overall sizes of the confidence regions rather than in their correlation patterns.

This result is consistent with the sensitivity and SNR analyses presented in Figs.~\ref{fig:sensitivity_LISA_TAIJI_VS} and \ref{fig:snr_ratio_VS}, where LISA-TAIJIp also achieves higher sensitivities and larger recovered SNRs for both vector and scalar polarizations. Unlike the TVS, TS, and TV scenarios, where the performance differences are closely related to the network's ability to mitigate tensor-related polarization degeneracies, the VS scenario appears to be dominated by the intrinsic sensitivity of the detector network. Consequently, the superior sensitivity of LISA-TAIJIp in the relevant frequency range translates directly into tighter Fisher constraints on the vector and scalar polarization parameters.

Taken together, Figs.~\ref{fig:corner_TVS}-\ref{fig:corner_VS} demonstrate a distinction between polarization separation and parameter estimation. The performance differences observed in the sensitivity and SNR studies should not be directly interpreted as equivalent differences in parameter estimation accuracy. The former quantify the capability of a detector network to separate different polarization sectors, whereas the latter quantify the precision with which the parameters of a prescribed signal model can be measured. Once the tensor, vector, and scalar components are explicitly incorporated into the signal model, a substantial fraction of the polarization information is already encoded in the parameterization itself, significantly reducing the impact of geometrical degeneracies.

The above conclusions should therefore be interpreted within the context of the adopted assumptions. If the SGWB spectrum deviates significantly from the assumed power-law form, or if a model-independent reconstruction of the stochastic background is pursued, the advantages provided by the parametric description would be reduced. In such situations, the intrinsic geometrical properties of the detector network become increasingly important, and the stronger polarization separation capability of the LISA-TAIJIm configuration is expected to play a more significant role. The Fisher forecasts presented here therefore complement, rather than contradict, the sensitivity and SNR analyses discussed in the previous section.

\section{Conclusions and discussion} \label{sec:conclusions}

Future space-based gravitational-wave detector networks will provide unprecedented opportunities for detecting SGWBs and probing their polarization content. In this work, we investigated the capabilities of two proposed LISA-TAIJI network configurations, LISA-TAIJIp and LISA-TAIJIm, for detecting and discriminating tensor, vector, and scalar polarization components of isotropic SGWBs. The two configurations share similar mission architectures but differ significantly in their relative constellation orientations, providing a useful framework for assessing the impact of network geometry on SGWB observations.

Using the ORFs of the two detector networks, we first evaluated their sensitivities and detection performances for SGWBs containing different combinations of polarization sectors. We considered TVS, TS, TV, and VS backgrounds within a model-independent component-separation framework. In all cases, the simultaneous presence of multiple polarization sectors degrades the sensitivity relative to the pure-polarization limit owing to correlations among the detector responses. The degradation becomes particularly severe in the TVS scenario, where tensor, vector, and scalar components must be separated simultaneously.

We found that the relative geometry of the detector network plays a crucial role in determining the severity of these polarization degeneracies. Compared with LISA-TAIJIp, the LISA-TAIJIm configuration exhibits substantially weaker correlations among tensor, vector, and scalar responses and therefore achieves significantly improved sensitivities and recovered SNRs in the TVS, TS, and TV scenarios. In the full TVS case, the recovered SNRs can differ by factors of a few between the two network configurations. In contrast, the VS scenario exhibits qualitatively different behavior: the performance difference between the two networks becomes much smaller, and LISA-TAIJIp can even outperform LISA-TAIJIm. This result indicates that the impact of detector geometry depends not only on the relative inclination between detector constellations but also on the specific polarization mixture present in the SGWB.

We further investigated the constraining capabilities of the two detector networks using Fisher-matrix forecasts for polarized SGWB models described by power-law spectra. For the TVS, TS, TV, and VS scenarios, the projected constraints on the polarization amplitudes and spectral indices were found to be broadly similar for the LISA-TAIJIp and LISA-TAIJIm configurations, despite the substantial differences observed in the sensitivity and SNR analyses. This behavior originates from the construction of the Fisher matrix itself. Since the derivatives are taken with respect to the parameters of individual polarization components, the tensor, vector, and scalar sectors are already treated as distinct contributions within the assumed signal model. Consequently, the Fisher analysis evaluates the precision with which the parameters of a prescribed polarized SGWB model can be measured, rather than the detector network's capability to separate different polarization sectors directly from the data. As a result, the strong polarization degeneracies that dominate the model-independent component-separation analysis have a much weaker impact on the forecasted parameter constraints.

Taken together, our results demonstrate that the relative orientation of space-based detector constellations is a key factor for model-independent SGWB polarization measurements. Networks with larger relative inclinations provide more linearly independent responses to different polarization sectors and therefore exhibit superior polarization-separation capabilities, particularly when tensor and non-tensorial modes coexist. At the same time, the dependence of parameter constraints on detector geometry can be substantially reduced when strong parametric assumptions about the SGWB spectrum are imposed. These findings highlight the complementary roles of model-independent polarization reconstruction and model-dependent parameter inference in future SGWB studies.

Several caveats should be noted. Throughout this work, we assumed isotropic SGWBs described by power-law spectra and considered only cross-correlations between the LISA and TAIJI missions. More realistic backgrounds may possess nontrivial spectral features structures that are not captured by the present framework. Future investigations incorporating nonparametric spectral reconstructions and Bayesian parameter inference will provide a more comprehensive assessment of the polarization-measurement capabilities of space-based detector networks.

This work was performed by using the python packages \textsf{numpy} \cite{harris2020array}, \textsf{scipy} \cite{2020SciPy-NMeth}, \textsf{pandas} \cite{pandas}, and the plots were made by utilizing \textsf{matplotlib} \cite{Hunter:2007ouj}, \textsf{GetDist} \cite{Lewis:2019xzd}, and \textsf{HEALPix} \cite{2005ApJ...622..759G,Zonca2019}.

\begin{acknowledgments}

G.W. was supported by the National Key R\&D Program of China under Grant No. 2021YFC2201903 and NSFC Grant No. 12575058. L.L. is supported by the National Natural Science Foundation of China (Grant No.~12505054, 12447101 and ~12433001) and the Fundamental Research Funds for the Central Universities.

\end{acknowledgments}

\bibliography{apsref}

\end{document}